\documentclass{emulateapj}

\pdfoutput=1

\usepackage{hyperref}
\usepackage{natbib}
\usepackage{color}
\usepackage{soul}
\usepackage{subfigure}
\soulregister\citet7
\soulregister\citep7
\soulregister\citealp7
\soulregister\footnote7
\soulregister\equation7
\soulregister\subsection7
\soulregister\ref7
\soulregister\se7
\soulregister\Fig7
\soulregister\Eq7
\soulregister\hi7
\soulregister\ha7
\soulregister\magarc7
\soulregister\kms7
\soulregister\solarm7
\usepackage{amsmath}
\usepackage{wasysym}
\usepackage{threeparttable}
\usepackage{booktabs}
\usepackage{multirow}
\usepackage{changepage}

\newcommand{\kms}{\ifmmode\,{\rm km}\,{\rm s}^{-1}\else km$\,$s$^{-1}$\fi}
\newcommand{\Rd}{\ifmmode\,R_{\rm d}\else $R_{\rm d}$\fi}
\newcommand{\be}{\begin{equation}}
\newcommand{\ee}{\end{equation}}
\newcommand\ltsima{$\; \buildrel < \over \sim \;$}
\newcommand\ltsim{\lower.5ex\hbox{\ltsima}}
\newcommand\gtsima{$\; \buildrel > \over \sim \;$}
\newcommand\gtsim{\lower.5ex\hbox{\gtsima}}
\newcommand{\solarm}{\ensuremath{M_\odot}}

\newcommand{\magarc}{\ifmmode {{{{\rm mag}~{\rm arcsec}}^{-2}}}
             \else {{{mag}$~${arcsec}$^{-2}$}}
             \fi}
             
\newcommand{\msol}{\mbox{\,$M_\odot$}}        % solar mass                                                       
\newcommand{\mstel}{\mbox{\,$M_{*}$}}        % stellar mass                                                     
\newcommand{\Cla}{\hyperlink{C14}{C14}}
\newcommand{\Clb}{\hyperlink{C17}{C17}}
\newcommand{\Ca}{\hyperlink{C07}{C07}}
\newcommand{\Esk}{\hyperlink{E12}{E12}}
\newcommand{\Me}{\hyperlink{M14}{M14}}
\newcommand{\Ken}{\hyperlink{K98}{K98}}

\def\Eq#1{Eq.~\ref{eq:#1}}
\def\eq#1{eq.~\ref{eq:#1}}
\def\Fig#1{Fig.~\ref{fig:#1}}
\def\Table#1{Table~\ref{#1}}
\def\sec#1{Sec.~\ref{sec:#1}}

\def \ion#1#2{#1{\footnotesize{#2}}\relax}

\newcommand{\ha}{H$\alpha$}
\def \hi{\ion{H}{I}}

\slugcomment{Submitted to Astrophysical Journal}
\begin{document}

\title{Unique Tracks Drive the Scatter of the Spatially-Resolved Star Formation Main Sequence}
\shorttitle{Scatter of the Star Formation Main Sequence}
\shortauthors{Hall et al.}

%\correspondingauthor{Christine Hall\altaffilmark{1}}
%\email{1cfh@queensu.ca}
\author{
Christine Hall\altaffilmark{1},%\footnote{\href{mailto:1cfh@queensu.ca}{1cfh@queensu.ca}},
St\'ephane Courteau\altaffilmark{1},
Thomas Jarrett \altaffilmark{2},
Michelle Cluver \altaffilmark{3,4},
Gerhardt Meurer \altaffilmark{5},
Claude Carignan \altaffilmark{2},
Fiona Audcent-Ross \altaffilmark{5}
}
\altaffiltext{1}{Department of Physics, Engineering Physics and Astronomy, Queen's University, Kingston, ON K7L 3N6, Canada}
\altaffiltext{2}{Department of Physics and Astronomy, University of Cape Town, Cape Town South Africa}
\altaffiltext{3}{Centre for Astrophysics and Supercomputing, Swinburne University of Technology, Hawthorn, VIC 3122, Australia}
\altaffiltext{4}{Department of Physics and Astronomy, University of the Western Cape, Robert Sobukwe Road, Bellville, 7535, South Africa}
\altaffiltext{5}{International Centre for Radio Astronomy Research, The University of Western Australia, 35 Stirling Highway, Crawley, WA 6009, Australia}

\begin{abstract}

The scatter of the spatially resolved star formation main sequence (SFMS) is investigated in order to reveal signatures about the processes of galaxy formation and evolution. We have assembled a sample of 355 nearby galaxies with spatially resolved H$\alpha$ and mid-infrared fluxes from the Survey for Ionized Neutral Gas in Galaxies and the \textit{Wide-field Infrared Survey Explorer}, respectively.  We examine the impact of various star formation rate (SFR) and stellar mass transformations on the SFMS. Ranging from 10$^{6}$ to 10$^{11.5}$\msol\ and derived from color to mass-to-light ratio methods for mid-infrared bands, the stellar masses are internally consistent within their range of applicability and inherent systematic errors; a constant mass-to-light ratio also yields representative stellar masses. The various SFR estimates show intrinsic differences and produce noticeable vertical shifts in the SFMS, depending on the timescales and physics encompassed by the corresponding tracer. SFR estimates appear to break down on physical scales below 500~pc.  We also examine the various sources of scatter in the spatially resolved SFMS and find morphology does not play a significant role. We identify three unique tracks across the SFMS by individual galaxies, delineated by a critical stellar mass density of log($\Sigma_{\mstel}$)$\sim$7.5. Below this scale, the SFMS shows no clear trend and is likely driven by local, stochastic internal processes.  Above this scale, all spatially resolved galaxies have comparable SFMS slopes but exhibit two different behaviors, resulting likely from the rate of mass accretion at the center of the galaxy.
\end{abstract}

%----------------------------------------------------------INTRODUCTION-----------------------------------------------------%
\section{Introduction}
\label{sec:intro}

Studies of the star formation main sequence (SFMS) of galaxies, the relation between star formation rates (SFRs) and stellar mass (\mstel), for galaxies throughout our universe have revealed that galaxy formation is an orderly process, and universal laws must govern their evolution throughout cosmological time and across many different environments. 
Global SFRs rise with cosmic time (\citealt{noeske07,whit14,salmon15,kelson16,tomczak16}), whilst only exhibiting a moderate increase in scatter (\citealt{kurc16,mitra17}).
Of interest in the study of galaxy scaling relations, like the SFMS, is the ability to separate and identify the processes that lead to their development. Broadly speaking, morphology and galaxy structure correlate strongly with the SFMS. Two well-known, distinct sequences of SFR emerge as typically gauged by \ha\ emission equivalent widths (\textit{EW}(\ha)), separating star forming (blue cloud) and quiescent (red sequence) galaxies (\citealt{noeske07, eales17, oemler17, pandya17}). Similarly, the degree of light concentration, or Sersic index, is thought to influence the scatter of the global SFMS (\citealt{wuyts11, whit15, brennan17}), though its induced shift across the SFMS is more gradual (continuous) than that characterized by the blue cloud and the red sequence.
The global environment in which a galaxy evolves also affects its star formation (SF) activity.  Indeed, while the relation between environment, from void galaxies to clusters to centrals to satellites, and SF activity in galaxies is non-trivial, there is general agreement that quenching or passive environments prevail in increasingly dense environment \citep{peng12,fossati15,beygu16,coil17}.  The processes of gas supply (infalls) and depletion (outflows and consumption) that enable SF may also be traced through the SFMS's apparent dependence on metallicity (\citealt{mannucci10, wuyts11, obreja14, telford16}), or the \hi\ gas fraction \citep{saintonge16}.

On large scales, the dark matter halo in which a galaxy is embedded also affects the SFMS through the stellar mass-halo mass (\mstel-M$_{\mathrm{halo}}$) relation \citep{gu16,garrison17}. Disentangling which of the stellar or halo mass drives this relation is however challenging \citep{kimm09}, especially since processes of different nature dominate in different stellar mass ranges \citep{williams10,perez13,beygu16,magdis16}, with an increasing degree of scatter at lower \mstel\ \citep{guo13,obreja14}. It remains unclear if the scatter increase at lower \mstel\ arises from enhanced stochasticity in lower mass environments, or simply results from the uncertainty in the transformations that are applied.
Other studies have also found a correlation

Most studies of the SFMS and its scatter have relied on integrated, global values of the SFR and \mstel\ over each galaxy, despite the fact that SF processes display a well-known local dependence, through the star formation, or Kennicutt-Schmidt (K-S), law\footnote{In its most common form, the Kennicutt-Schmidt relation states that star formation densities scale with gas surface densities by $\Sigma_{\mathrm{SFR}}\propto \Sigma_{gas}^{N}$, where $N$=1-2 is a constant \citep{schmidt59,kennicutt98b,genzel13}.}. It follows that the gas fraction and surface density should influence the SFMS \citep{tacconi13,saintonge16}. Since this relationship arises on small scales, one must therefore examine the spatially resolved SFMS, and determine whether the connection between the star formation law and the SFMS holds in progressively smaller regions.
Several studies have recently capitalized on the newly available spatially resolved data, probing a local scale SFMS (see \citealt{perez13,wuyts13,hemmati14,cano16,gonzalez16,magdis16,abdurro17,gonzalez17,marag17,wang17,ellison18}). 
Such preliminary spatially resolved studies have demonstrated trends of comparable amplitude between the global and local SFMS properties such as their slope, zero-points, and scatter, whilst highlighting that variations in the SFMS on local scale do exist. 
The morphological dependence of the global SFMS shows late-type galaxies with the highest SFRs defining an upper boundary in the local SFMS, with progressively earlier systems moving across and below the local sequence \citep{gonzalez16,abdurro17,gonzalez17}.
Intriguingly, the global morphology seems to dictate local structural trends within the galaxy by establishing the vertical offset from the SFMS fit, whilst maintaining minimal scatter, as measurements from galaxies of the set morphology run along the same SFMS slope.

One of the more powerful observations afforded by spatially resolved studies may arise from examining the shapes of specific SFR (sSFR = SFR/\mstel) radial profiles. These profiles are tell-tales of quenching behavior across the galaxy, presenting strong motivation for both inside-out quenching (\citealt{forbes14,gonzalez16,tacchella16b,belfiore17,ellison18}) and outside-in quenching \citep{schaefer17,medling18}, and possibly suggesting a shift between inside-out vs. outside-in quenching behavior dictated by total \mstel\ \citep{kimm09,perez13,liu18}.

Observational studies of the SFMS can also be contrasted with models and simulations at both global and local scales \citep{dutton10,somerville15,lagos16,rodriguez16,tacchella16a,tacchella16b,brennan17,pandya17,matthee18}.
Simulations can highlight the quenching trends and physical mechanisms that may drive the SFMS over time \citep{tacchella16b, pandya17, wang17};
for instance, \cite{tacchella16a} surmises that the scatter of the SFMS reflects the oscillation of a galaxy about the SFMS through different evolutionary phases. These phases may ultimately give rise to the separation between star forming and quiescent sequences \citep{pandya17}.  Careful data-model comparisons can indeed complete our understanding of the SFMS scatter.

This study recognizes that spatially resolved investigations of galaxies are the key to understanding the fundamental underpinnings of the SFMS.  We wish to identify basic drivers of the SFMS as determined empirically.  We can then expand upon the realm of empirical investigations by exploring universal trends on the smallest possible galactic scales. To this end, we wish to examine a comprehensive suite of radial profiles (mass-to-light ratios, stellar masses, SFRs, specific SFRs, and SFR or stellar mass densities) out to larger radial values than most current studies of the global and spatially resolved (local) SFMS. We can also correlate the scatter of the SFMS with such parameters, and establish connections with numerical models and theoretical investigations accordingly.

The SFMS is characterized by the linear correlation, $\log{(\mathrm{SFR})} = a\log{(\mstel)} +b$, and great efforts have been invested to measure the variations and dependence of $a$ and $b$. \cite{speagle14} provide a comprehensive review of SFMS studies out to $z\approx6$ with corresponding SFMS fit results for $a$ and $b$ and the relation scatter over the last decade; slope values ($a$) ranging from 0.05 to 1, and zero point values ($b$) ranging from -9.6 to 0.8 are  reported. A turnover is commonly observed above $\log(\mstel/\msol) \approx 10.5$ at low redshift, suggesting that the slope ($a$) decreases at this critical mass \citep{lee15,tomczak16}; though there remains discussion whether this is intrinsic to the SFMS, or arises from differences between SF and quiescent galaxies \citep{pearson18}.
The substantial variation in $a$ and $b$  have largely been attributed to the evolution in SFRs over cosmic time; a definitive understanding of the SFMS scatter is however still lacking (even though the path that a galaxy follows in the SFMS is clearly a reflection of its mass accretion history and its ability to turn gas into stars across various environments).

In order to study the scatter of the global and local SFMS, our study takes advantage of our own spatially resolved stellar mass maps and SFRs for an ensemble of over 350 low-redshift galaxies, allowing us to probe local SFR and \mstel\ conditions across the galaxy. The variations of the slopes ($a$), zero-point ($b$), and scatter ($\sigma$) of the global and local SFMS should yield a clearer appreciation of the processes driving SF on all relevant scales.

The outcome of such spatially resolved SFMS investigations lies critically in the choice of SFR and \mstel\ transformations. SFR estimates are made by exploiting spatially resolved \ha\ flux profiles from the Survey for Ionized Neutral Gas in Galaxies (SINGG, \citealt{meurer06}) and 12$\mu$m and 23$\mu$m band profiles from the \textit{Wide-field Infrared Survey Explorer} (\textit{WISE}, \citealt{wright10}). Stellar masses (\mstel), which are largely controlled by the integrated SF, can be recovered from the spatially resolved \textit{WISE} 3.4$\mu$m and 4.6$\mu$m band profiles. 

The data and methods to extract SFRs and {\mstel}'s are described in \sec{data} \& \ref{sec:methods}. Special attention is paid to the dependence of our results on the choice and consistency of those methods. The influence of the choice of spatial resolution scale is discussed in \sec{reso}. The global and local SFMS is reported in \sec{results} and the scatter is analyzed in \sec{scatter} \& \ref{sec:radial}. Conclusions are summarized in \sec{summary}.

%----------------------------------------------------------DATA-----------------------------------------------------%
\section{Data}
\label{sec:data}

Our study takes advantage of large, spatially resolved, complementary data sets for two extensive surveys of galaxies: the Survey for Ionization in Neutral Gas Galaxies (SINGG) and \textit{Wide-field Infrared Survey Explorer} (\textit{WISE}).

The SINGG targets were selected from the \hi\ Parkes All Sky Survey (HIPASS), which maps \hi\ 21cm line emission, representing regions that likely fuel SF. Of the 468 HIPASS targets, 289 were selected to also be mapped by SINGG. From these targets, 463 distinct galaxies were identified in \ha\ by these observations, with many HIPASS targets consisting of multiple galaxies \citep{meurer06}. 
SINGG provides resolved \ha\ and $R$-band maps of these nearby star forming galaxies \citep{meurer06}. 
These southern hemisphere galaxies have an average redshift of $z\sim0.01$, with \hi\ masses ranging from $7.0 < \log_{10}(\mstel/M_{\odot}) < 11.0$ \citep{hanish06}. The subselection from HIPASS favors the nearest galaxies at any given \hi\ mass. 
Capitalizing on this subselection alleviates usual optical biases (ex. total luminosity, surface brightness, or Hubble type; \citealt{meurer06}). A broad range of star-forming environments is thus enabled. The sample has no galaxy inclination cut either. The \ha\ emission line flux is representative of ionized gas surrounding newly formed, massive stars, and is used as an appropriate tracer for SFR. In addition to galacto-centric radial H$\alpha$ flux growth profiles and surface brightness profiles, SINGG provides 
complementary total \hi\ masses.
 
The \textit{WISE} data set provides spatially resolved photometry in the \textit{W}1 (3.4$\mu$m), \textit{W}2 (4.6$\mu$m), \textit{W}3 (12$\mu$m), and \textit{W}4 (23$\mu$m) bands for galaxies across the whole sky, and overlaps with 355 SINGG galaxies \citep{jarrett13,brown14}. The angular resolution of these bands are 5$^{\prime\prime}$.9, 6$^{\prime\prime}$.5, 7$^{\prime\prime}$.0, and 12$^{\prime\prime}$.4, respectively \citep{jarrett12}. 
The \textit{W}1 and \textit{W}2 bands are sensitive to evolved stellar populations and hence enable the calculation of reliable stellar masses; while the \textit{W}3 and \textit{W}4 bands, which have been corrected to remove stellar emission, 
sample re-radiated light from dust warmed by hot young stars and act as alternative SFR tracers. \textit{WISE} provides radial flux growth profiles, surface brightness profiles, concentration indices, as well as effective radii and surface brightness for each band. The effective radius, $R_{\rm{eff}}$, is the radius within which half the total light of the galaxy is contained. The effective surface brightness, SB$_{\rm{eff}}$, is then the surface brightness at this radius.

These resolved data offer a unique perspective on the global and spatially resolved (local) SFMS. The \ha\ and \textit{WISE} radial profiles of each individual galaxy have been rebinned to a matching scale, in order to directly compare each band.
\textit{WISE} resolution can probe regions as small as 90-200~pc across for our closest targets and up to a few kpc for our most distant systems. 
One must consider whether these spatial scales are a proper match to the transformations being applied and to our theoretical expectations for the correlation between SFR and stellar mass (given that the latter is the integral of the former). The impact and motivation of binning our profiles on various physically-meaningful scales is explored in \sec{reso}. Ultimately, our profiles will be rebinned to a spatial scale of 500~pc. With 355 galaxies, this produces a statistically robust local main sequence with nearly 6,000 data points.

A tremendous advantage also afforded by these data sets is their large radial extent, reaching out to almost 10 effective radii in some galaxies, unlike most existing surveys which often only probe out to 1.5~$R_{\mathrm{eff}}$. Our data allow us to trace SFR and \mstel\ conditions in commonly unexplored outskirts of a galaxy, reaching low SFRs and surface brightnesses, and encompassing the full variations of SFR and \mstel\ across a galaxy.

%----------------------------------------------------------METHODS-----------------------------------------------------%
\section{A Review of Transformation Methods}
\label{sec:methods}

\subsection{Star Formation Rates}
\label{sec:sfrs}

%**********************************************************
%%Figure: compare SFRs
%
\begin{figure*}[t!]
\centering
\includegraphics[width=1.0\textwidth]{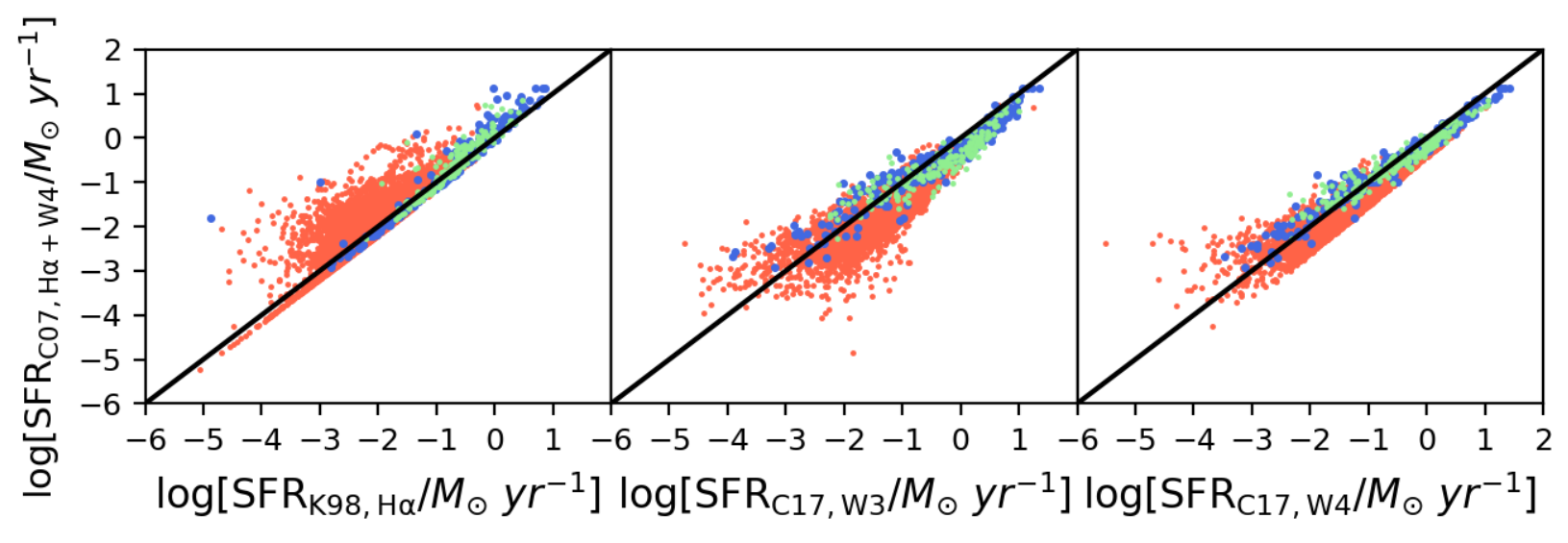}
\caption{SFRs inferred from the  \cite{kennicutt98b} (\ha), \cite{calzetti07} (\ha\ $+$ \textit{W}4) and \cite{cluver17} (\textit{W}3 and \textit{W}4) transformations. Local (ie. spatially resolved) values throughout each galaxy are in orange, integrated values out to $R_{\mathrm{eff},W1}$ are in green, and total integrated galactic values are in blue. The one-to-one line overlaid in black.}
\label{fig:SFRs}
\end{figure*}
%
%**********************************************************

There exists a multitude of transformations to infer SFRs in galaxies. These are typically based on UV spectrum slopes, H$\alpha$ and far-infrared (IR) fluxes, and more (see \citealt{kennicutt98b,calzetti13,davies16} for a review). These derivations can generally be traced back to motivation by the star formation (or K-S) law, which describes the SFR dependency on the total gas density and photo-reionization models \citep{schmidt59,kennicutt98a,kennicutt98b}, though varying ranges of SFRs can be obtained. This variation stems from the use of different wavebands, stellar population models, initial mass functions (IMFs), dust absorption models, etc. applied to each calibration; e.g. \cite{kennicutt83}. In principle, one single SFR ought to characterize a particular region over a given timescale; the source of SFR differences is also examined below.

The most common transformation is that of \cite{kennicutt98b} (hereafter \Ken), where the \ha~ emission line flux becomes a tracer of instantaneous star formation. This emission arises from ionized hydrogen surrounding a newly formed star, re-emitting the light produced by the young star. This proxy for SF applies to short timescales of $\lesssim$20~Myr, as only massive (O and B) stars with those ages can heat the gas to suitable temperatures (\Ken, \citealt{davies16}). 
Assuming a Salpeter IMF, the SFR conversion is as follows;
\be\label{eq:KSFR}
\mathrm{SFR}_{\mathrm{K98,H\alpha}} (\msol~\mathrm{yr^{-1}}) = 7.9\times 10^{-42}~L_{H\alpha} (\mathrm{erg~s^{-1}}).
\ee
The largest limitation of this method is accounting for extinction in the \ha~emission, which is specific to the environment of study \Ken. 

Young stars also emit copious amounts of ultraviolet photons, which can be absorbed by dust and re-emitted at mid-infrared (MIR) wavelengths, making that light another reliable tracer of SF (\citealt{cluver14}, hereafter \Cla). The process of light absorption and re-radiation is slow and MIR fluxes typically trace SF over longer timescales of $\sim$100 Myr. \cite{cluver17} (hereafter \Clb) have developed two SFR conversions motivated by a Kroupa IMF for the \textit{WISE} \textit{W}3 (12$\mu$m) and \textit{W}4 (23$\mu$m) bands;
\be\label{eq:ClSFR1}
\log(SFR_{\mathrm{C17},W3}/\msol~\mathrm{yr^{-1}}) = 0.889~\mathrm{log}(L_{W3}/L_{\odot}) - 7.76,
\ee
\be\label{eq:ClSFR2}
\log(SFR_{\mathrm{C17},W4}/\msol~\mathrm{yr^{-1}}) = 0.915~\log(L_{W4}/L_{\odot}) - 8.01,
\ee
\noindent If the light is not fully re-radiated, these formulae may actually underestimate the true SFR. 
The \textit{W}4 conversion (23$\mu$m; \Eq{ClSFR2}) may be preferable over the \textit{W}3 (12$\mu$m; \Eq{ClSFR1})) conversion, as the latter lies within the regime of emission from excited polycyclic aromatic hydrocarbon (PAH) molecules \citep{jarrett13}, and therefore have a more complex mapping to SFR.  
The \textit{W}4 transformation, which lies beyond the bulk of PAH emission, is therefore favored here. \cite{rosario16} compared the corresponding \textit{W}4 SFRs to those derived from optical tracers (ie. \ha) in star-forming galaxies and found comparable agreement. Caution is offered for survey galaxy selections motivated by the \textit{W}4 band emission which may bias the analysis towards warm dust regions, and valuable information from cold dust regions would be lost. Fortunately, our target selection was motivated by SINGG, and therefore HIPASS, which is sensitive to \hi~ emission and unbiased by warm dust.

Yet another SFR transformation by \cite{calzetti07} (hereafter \Ca) capitalizes on both \ha\ and MIPS 24$\mu$m to account for dust obscured and unobscured SF environments, in order to produce robust SFR estimates over short and long timescales.  Therefore, this calibration should hold over all scales across a galaxy.
The hybrid conversion adopts a Kroupa IMF from Starburst99 and is as follows;
\be\label{eq:CSFR}
\begin{split}
\mathrm{SFR}_{\mathrm{C07},H\alpha+24\mu m} (\msol~\mathrm{yr^{-1}}) &= 5.3\times 10^{-42}~[L_{H\alpha} (\mathrm{erg~s^{-1}}) \\
& + 0.031*L_{\nu,24\mathrm{\mu m}} (\mathrm{erg~s^{-1}~Hz^{-1}})].
\end{split}
\ee % this eq'n is calibrated to MIPS24
\noindent This conversion is calibrated to the MIPS 24$\mu$m band, which differs slightly from the \textit{W}4 band; see Fig.~2 of \cite{brown14}. Since, these band responses are reasonably consistent, this conversion should be applicable to \textit{W}4 luminosities, although the differences should be in the sense that SFRs will be slightly underestimated using the \textit{W}4 band. Fortunately, the overall \textit{W}4 contribution in this conversion is small, so the impact is arguably negligible.

\cite{calzetti13} present a thorough review and comparison of the many SFR indicators available, examining their motivating physics, unique advantages, and applicable scales. At large scales, single emission line tracers are most reliable under low attenuation conditions. However, as dust contributions increase, as is common in SF galaxies, then a hybrid combination of \ha\ and 24$\mu$m compensates for any attenuated line emission, by accounting for re-radiated dust emission at 24$\mu$m (see also, \citealt{lee13}).  On increasingly small scales and therefore regions of lower SFR, such indirect tracers become 
increasingly vulnerable to regions without SF that still emit such light. Across all SFR tracers, 
the generalization of underlying stellar populations and IMFs that occurs on large scales is likely no longer appropriate. This highlights the complexity and increased vulnerability of tracers at the local, resolved scale.  Overall, while each tracer has its own strengths or weaknesses, especially at low SFR regimes, emission line tracers may be the least vulnerable to underlying stellar population assumptions due to the short SF timescales \cite{calzetti13}.

\Fig{SFRs} compares SFR transformations for both local (orange) and integrated (blue and green) measurements. While global values display some scatter, it is considerably smaller than for local measurements. Innate scatter is expected by the nature of each transformation and the specific SF conditions and timescales of their corresponding tracer represents. However, the increased scatter from global to local values is significant, and must be understood.
The conversions by \Ca\ (\eq{CSFR}) and \Clb\ (\textit{W}4, \eq{ClSFR2}) produce the most consistent SFRs, while the former transformation has the highest overall agreement among all SFR transformations. This could be due to the fact that \ha\ and \textit{W}4 are also in the \Ken\ and \Clb, though despite the dominant contribution of \ha\ in \Ca, there is greater consistency with the \textit{W}4 \Clb\ conversion.
The above calibrations, and their impact on the SFMS, are compared in \sec{results} in order to isolate the variations in SFR estimates from different tracers.  The \Ca\ transformation (\eq{CSFR}) is favored in this analysis since it utilizes both \ha, which remains stable across varying stellar populations, and \textit{W}4 emission, which accounts for dust attenuation effects.

\cite{catalan15} and \cite{davies16} also compare different SFR tracers and calibrated transformations. Their calibrations are constructed from CALIFA (Calar Alto Legacy Integral Field sepctroscopy Array) survey and GAMA (Galaxy and Mass Assembly) measurements, respectively. \cite{catalan15} calibrated \ha, UV and TIR based transformations, including hybrid combinations of these bands. Using \ha\ SFRs as the standard to calibrate to, the resulting SFRs from UV and TIR luminosities (or hybrid combinations) are matched to the standard with minimized scatter. 
The \cite{davies16} analysis accounted for \ha, \textit{W}3 and \textit{W}4 bands as SFR tracers, along with [OII] emission, near and far ultraviolet, u-band, 100$\mu$m, infrared, and spectral energy distribution fitting. Unique to this study, they examined the consistency of the resulting SFR estimates from each method and their effect on the SFMS. 
Assuming that SFRs should be the same regardless of the transformation (though this should be cautioned against, as SFR are dependent on the representative timescale of that tracer), they use SFR calculation discrepancies and SFMS fits to propose revised SFR calibrations for each tracer that ultimately produces SFRs and a SFMS with highest fidelity. The reported hybrid calibration with \ha\ and \textit{W}4 by \cite{catalan15} is consistent with the \Ca\ calibration (\eq{CSFR}); the revised calibrations by \cite{davies16} are unique.  In \sec{results}, the \cite{davies16} calibrations are applied to our data and interpreted in the context of measurements external to the GAMA survey.

\subsection{Stellar Masses}
\label{sec:stellm}

%**********************************************************
%%Figure: compare M/Ls
%
\begin{figure}[b]
\centering
\includegraphics[width=0.45\textwidth]{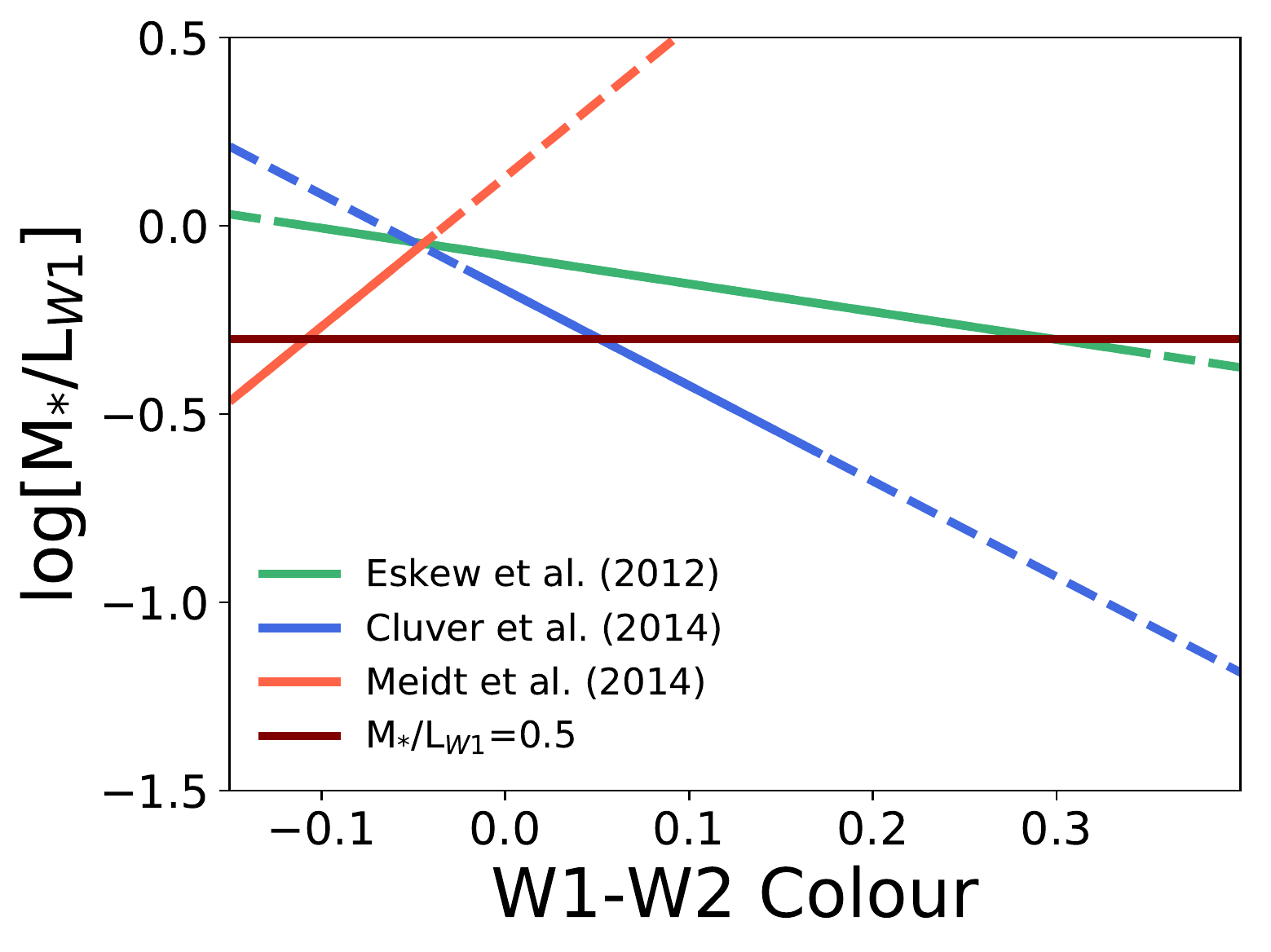}
\caption{Range of mass-to-light ratios ($M_{*}/L_{W1}$'s) produced by the stellar transformations by \cite{eskew12} in green, \cite{cluver14} in blue, and \cite{meidt14} in orange. Solid lines represent the applicable color range for each transformation.  Dashed lines represent regions where the transformation is no longer physical. The maroon line represents the constant $M_{*}/L_{W1}=0.5$.}
\label{fig:MLs}
\end{figure}
%
%**********************************************************
%Figure: compare M*s
%
\begin{figure*}[t!]
\centering
\includegraphics[width=1.0\textwidth]{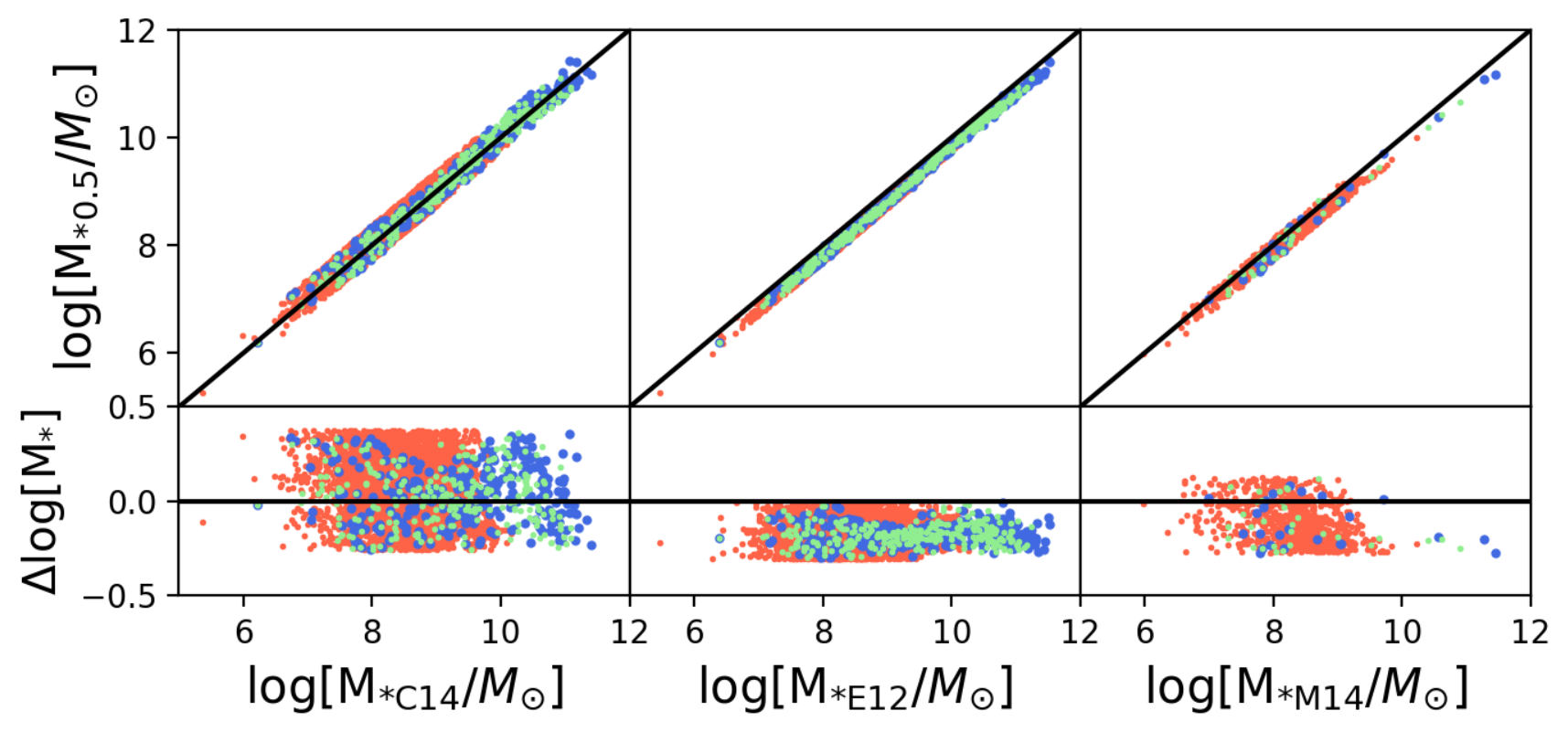}
\caption{Upper panel: Stellar masses from the \cite{eskew12}, \cite{cluver14}, and \cite{meidt14} transformations against stellar masses from a constant $M_{*}/L_{W1}=0.5$.
Lower panel: Difference in stellar masses from the \cite{eskew12}, \cite{cluver14} and \cite{meidt14} transformations compared to the constant $M_{*}/L_{W1}=0.5$ method. Local values throughout each galaxy are in orange, integrated values out to $R_{\mathrm{eff},W1}$ are in green, and total integrated galactic values are in blue. The zero difference, or one-to-one, line is overlaid in black.}
\label{fig:mstars}
\end{figure*}
%
%**********************************************************

Stellar masses can be derived through multiple methods; these include color to mass-to-light ratio relations (CMLRs), multiband SED fitting, or straight band flux to stellar mass transformations. Each method takes advantage of specific wavelengths, and this study capitalizes on the availability of the \textit{WISE} 3.4$\mu$m and 4.6$\mu$m bands. \cite{jarrett13} advocate the advantage of near infrared light since low-mass, evolved stellar populations dominate the emission in that wavelength range and comprise most of a galaxy's stellar mass.  With the appropriate initial mass functions and $M_{*}/L$ range, the total stellar mass in a region can be determined from this IR light.

CMLRs have been widely studied, especially for optical bands \citep{taylor11,roediger15,zhang17}. 
Optical $M_{*}/L$ rise with color and dust emission effects are feeble in this wavelength regime. Both \cite{taylor11} and \cite{roediger15} have shown that optical colors result in strongly linear relations since degeneracy from dust, age or metallicity scatter along the relation, rather than away it. They use SED fitting (though with different algorithms) to derive stellar masses and extract CMLRs. 
\cite{roediger15} emphasize, however, that the reliability of $M_{*}$ estimates improves with the number of bands utilized, thus favoring SED fitting.

An increase in dust emission in SF galaxies translates to a reddening of MIR colors, with $W1-W2 > 0$.  This reddening is unlike that for optical colors from low mass stars (which would correspond to increasing $M_{*}/L_{W1}$), whereas red MIR colors lower $M_{*}/L_{W1}$ as the light emission is largely dominated by dust \citep{quere15}. One must distinguish whether the detected emission has been corrected for dust effects or not, in order to apply the appropriate calibration to stellar mass. MIR CMLRs are highly sensitive to dust and the presence of dust changes mechanisms driving the relation between the MIR color and $M_{*}/L_{W1}$. Such trends are demonstrated in \Fig{MLs}. If $W1-W2>0$, dust is typically present. Although \cite{wright10} and \Cla\ show in their Figures 12 and 5 (respectively) that the range of $W1-W2$ colors corresponding to spiral galaxies is roughly between -0.2 and 0.6; values exceeding $W1-W2=0.3$ represent AGNs or starbursts and should be avoided (\citealt{stern12}, \Cla).

These considerations are critical in seeking accurate stellar masses from the \textit{WISE} (MIR) data. Before selecting a stellar mass transformation for this analysis, the methods utilizing the 3.4$\mu$m and 4.6$\mu$m bands, we first compare the methods that apply for the 3.4$\mu$m and 4.6$\mu$m bands. We enforce that each transformation caters to different galactic environments (\Esk, \Cla, \Me) and note that these CMLR conversions can be applied only for a specific $W1-W2$ color range. 

The stellar mass conversion by \Cla\ improves upon the preliminary \textit{WISE} conversion by \cite{jarrett13}. The original transformation was derived using a Charbrier-type IMF that takes into account general models of star formation histories, stellar population age and composition, dust content, and AGN activity.  Improvements by \Cla\ use in-depth calibrations of GAMA stellar mass estimates by \cite{taylor11} for over 110,000 low redshift ($z$) galaxies.  It represents a robust sampling of galactic environments, similar to our study. The conversion is meant to apply to ``normal'' galaxies and avoids warmer regions that may contain AGN or starbursting environments where the relation breaks down (beyond $W1-W2=0.3$). Therefore strict color limits of $-0.02\leq W1-W2 \leq0.17$ have been enforced. The proposed relation is:
\be\label{eq:CMstar}
\log(\mstel/L_{W1})_{\mathrm{C14}} = -0.17 - 2.54\times (W1-W2).
\ee
\noindent where \mstel\ is recovered from the resulting \mstel/L$_{W1}$ by multiplying by the corresponding luminosity: $L_{W1} (L_{\odot}) = 10^{-0.4(M_{W1} - M_{\odot})}$, where $M$ is an absolute magnitude.

\cite{eskew12} (hereafter \Esk) proposed an alternate form of a \mstel\ conversion, directly calculated from \textit{W}1 and \textit{W}2 band fluxes. This transformation was derived by calibrating 3.6 and 4.5 $\mu$m fluxes to previously existing stellar maps of the Large Magellanic Cloud (LMC). The stellar mass maps by \Esk\ were derived from the detailed star formation history (SFH) maps produced by \cite{harris09}, who adopted a Salpeter IMF. With known SFHs, the uncertainty on resulting \mstel\ is greatly reduced and strengthens the calibration.

Additionally, with the conversion being calibrated to spatially resolved stellar maps, it can be appropriately applied to the resolved regions of our spiral targets. Like the \Cla\ relation, it has been calibrated over a wide range of environments, albeit increasing uncertainty in strongly SF (warmer) environments. The transformation is given as:
\be\label{eq:EMstar}
M^{*}_{\mathrm{E12}}(\mathrm{M_{\odot}}) = 10^{5.65}~F^{2.85}_{W1}(\mathrm{Jy})~F^{-1.85}_{W2}(\mathrm{Jy})~\bigg[\frac{D(\mathrm{Mpc})}{0.05}\bigg]^{2}.
\ee
\noindent In order to compare this transformation with that of \Cla\ (\eq{CMstar}),  we have recast \eq{EMstar} using a similar notation as \eq{CMstar} (see App.~B):
\be\label{eq:EMstar2}
\log(\mstel/L_{W1})_{\mathrm{E12}} = -0.08 - 0.74\times (W1-W2).
\ee
\noindent The appropriate color range for their transformation is not explicitly specified, though the limits can easily be calculated from their Figures 3 \& 4. We estimate their limits to be $-0.12\leq W1-W2\leq0.34$. This is much broader than those reported by \Cla, as seen in \Fig{MLs}; however the resulting $M_{*}/L_{W1}$ range remains physical. These limits just breach into the range of colors that may result from AGN and strongly SF environments ($W1-W2>0.3$) that \cite{stern12} and \Cla\ caution against.
Although \Esk\ did not derive this conversion specifically for the \textit{WISE} bands, we argue that the 3.6 and 4.5 $\mu$m \textit{Spitzer} bands are reasonably consistent with the 3.4 and 4.6 $\mu$m \textit{WISE} bands.  This calibration may thus be applied to \textit{WISE}, as the respective bands trace the same behavior.
A disadvantage of this calibration is the low-metallicity of to the LMC; thus the conversion mostly applies to similarly low-metallicity regions.

\cite{meidt14} (hereafter \Me) produced an alternate conversion using spatial maps of 2300 galaxies from the \textit{Spitzer} Survey of Stellar Structure in Nearby Galaxies (S$^{4}$G). These maps preserve the structural information that is necessary in our study of the spatially resolved SFMS. Their derivation involves fitting [3.6]-[4.5]$\mu$m colors to previously calculated $M_{*}/L_{3.6}$'s that assume a Chabrier IMF, with a focus on resolving the age-metallicity degeneracy, which is not accounted for in the \Esk\ study, yielding the relation:
\be\label{eq:MMstar}
\log(\mstel/L_{W1})_{\mathrm{M14}} = 0.13 + 3.98\times (W1-W2).
\ee
This calibration, however, requires first removing the contribution of ``contaminants'' from the detected emission, in order to isolate the old stellar populations. Such contaminants in the [3.6]-[4.5]$\mu$m range from hot dust emission, asymptotic giant branch stars to red supergiants. \Me\ outlines the procedure to eliminate such non-stellar emission by their ``Independent Component Analysis'' (ICA).  To separate purely old stellar populations further, they restrict their study to regions with blue MIR colors, specifically $-0.15<[3.6]-[4.5]<-0.02$. 
This is consistent with the $W1-W2<0$ regime discussed above for dust-corrected, old stellar populations. 
This method therefore applies to data sets where dust contributions to MIR emission have been eliminated.

%**********************************************************
%%Figure: Binning
%
\begin{figure*}[t!]
\centering
\includegraphics[width=0.7\textwidth]{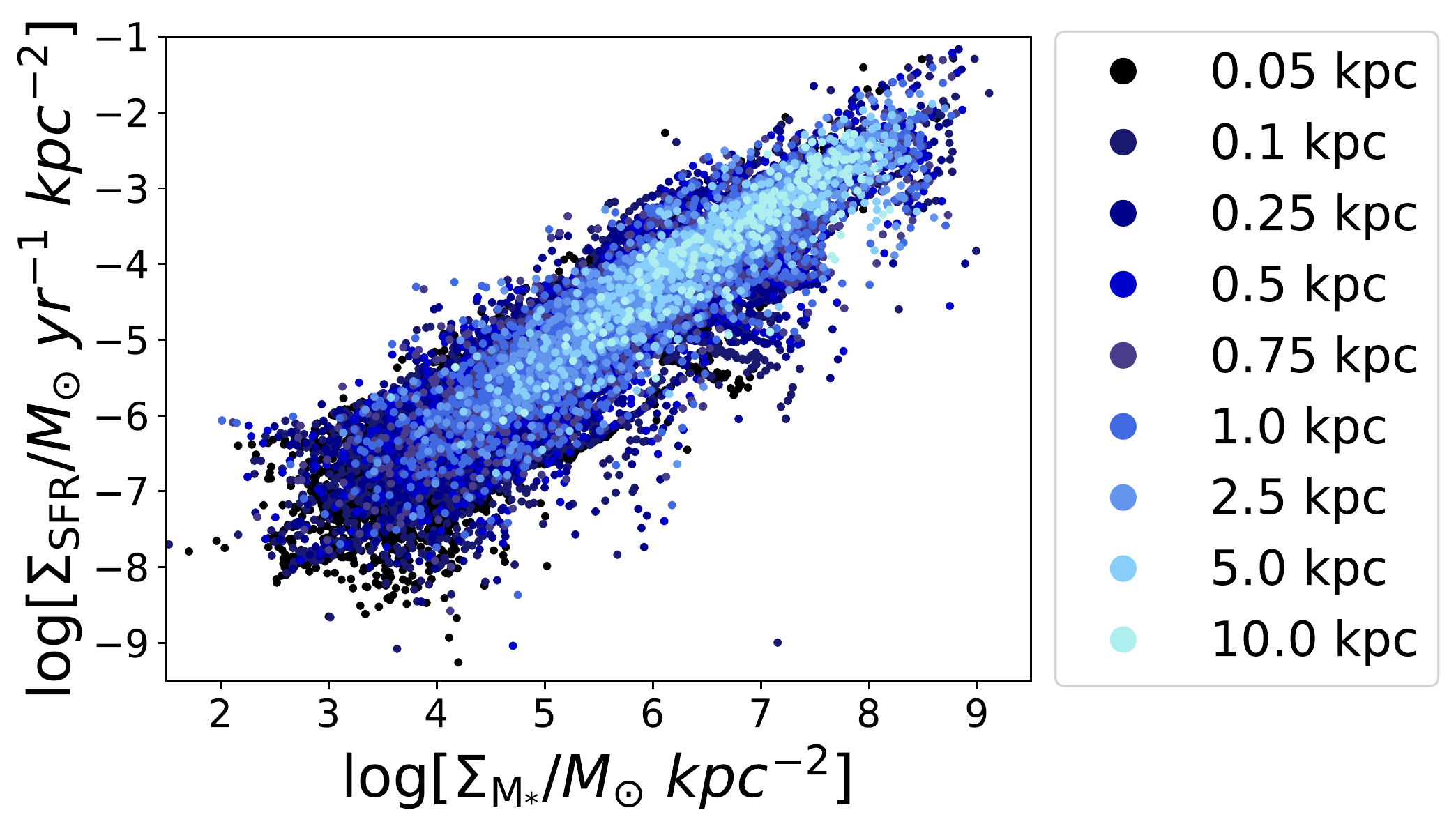}
\caption{The SFMS with $\Sigma_{\mstel}$ and $\Sigma_{\mathrm{SFR}}$ measurements at various spatial scales. Spatial scales are binned at 50~pc to 10~kpc resolution. The slopes and zero-points corresponding to local SFMS relations measured on different scales hardly change.}
% , as indicated in the legend.}
\label{fig:binning}
\end{figure*}
%
%**********************************************************

\Me\ also demonstrate the vulnerability of stellar masses to the adopted IMF. For their preferred Chabrier IMF, they report an average constant $M_{*}/L_{W1}=0.6$; for a Salpeter IMF, $M_{*}/L_{W1}$ nearly doubles (1.07). 
Fortunately, the range of $M_{*}/L_{W1}$ for each transformation examined here is small and averages about 0.5. \Fig{MLs} shows CMLRs based on both Chabrier (\Cla\ and \Me) and Salpeter (\Esk) IMFs; the resulting ranges in $M_{*}/L_{W1}$ show overall consistency.

Numerous studies have advocated for the characterization of a given galaxy stellar population composite by a constant $M_{*}/L$ whose uncertainty encompasses any variations possibly accounted for by CMLRs. 
The commonly proposed range, $M_{*}/L_{W1}=0.45-0.6$ (\Esk, \Me, \citealt{norris14,mcgaugh15,kettlety18, ponom18}). 
For instance, \cite{kettlety18} compared stellar masses from SED fits and CMLRs with those derived from a constant $M_{*}/L_{W1}$, and found matching results. The constant value of $M_{*}/L_{W1}=0.5$ shown in \Fig{MLs} is clearly a good match to the applicable color range and consistent within the uncertainty associated with stellar mass approximations.

The upper panel of \Fig{mstars} demonstrates the consistency of various $M_{*}/L$ methods, despite the varying MIR color ranges and $M_{*}/L_{W1}$ trends (\Fig{MLs}) amongst common transformations.  Here, stellar masses are compared for regions where the MIR color is appropriate for both \mstel\ transformations. Although the scatter about the one-to-one line (in black) is non-negligible, it is quite minimal compared to the uncertainty associated with such transformations.
This comparison of stellar mass estimates is similar to that of \cite{drory04}, and we observe a similar degree of scatter.

Interestingly, the increased discrepancies that arise in SFR estimates when calculating resolved local values versus total galactic values (\Fig{SFRs}), do not hold for stellar mass estimates. \Fig{mstars}, which compares \mstel\ values from each method, demonstrates that both total integrated \mstel\ values (in blue), \mstel\ values integrated out to $R_{\mathrm{eff}}$ (in green) and resolved local \mstel\ values (in orange) show a tight correlation about the one-to-one line, with no increased scatter in local values. The constant scatter for local and global values is emphasized in the lower panel of \Fig{mstars}, suggesting that the increased scatter in the local SFMS (\sec{local}) would be dominated by scatter in SFRs or local processes rather than from stellar masses.
The consistency amongst stellar masses integrated out to $R_{\mathrm{eff}}$ and total integrated values demonstrates that, despite a slight shift between those two, there is no significant growth in \mstel\ beyond $R_{\mathrm{eff}}$. Therefore global \mstel\ measurements are likely dominated by the inner regions of a galaxy.

In this latter figure, sharp colors limits are imposed in accordance with the \mstel\ transformation.
Within those regions, the stellar mass estimation methods are clearly consistent.
However, beyond these color limits, the physical environments, involving starbursts and AGN activity, are less well constrained and so are the \mstel\ estimates.  Such regions are excluded from our study, and hence sharp residual limits are seen in \Fig{mstars}.

Since we do not correct for dust emission, dust free and dust polluted regions could potentially be blended.  The $W1-W2$ colors in each region may serve as an indicator for such emission; if $W1-W2<0$, the dust contribution is likely minimal and the \Me\ transformation should be applied. Conversely, if $W1-W2>0$, then dust is likely contributing some of the emission and either \Esk\ or \Cla\ can be applied.

Such uncertainty justifies the use of a constant $M_{*}/L_{W1}$ ratio (\Me, \citealt{norris14,lelli16,kettlety18,ponom18}). \Fig{mstars} highlights that a constant $M_{*}/L_{W1}$ produces fairly consistent stellar masses with all three transformations and appropriate MIR color range. Though stellar masses may be slightly underestimated compared to the \Esk\ and \Me\ transformations, the resulting values are nonetheless consistent with those estimated in other studies, such as \cite{gonzalez16} or \cite{ellison18} (see \Fig{theSFMS}). This study traces considerably more low mass systems than CALIFA or MaNGA, which are limited to log(M$_{*,tot}/\msol)>9$;  the upper limits in stellar mass are consistent \citep{walcher14,ellison18}.
Additionally, the constant $M_{*}/L_{W1}$ is not limited by color range. For this reason, we adopt a constant $M_{*}/L_{W1}=0.5$ for our \mstel\ estimates across the full range of colors.

We have tested for other conversions and obtained statistically consistent SFMS results for the various stellar masses and SFRs (compare Table~\ref{fitstable} and \ref{app:fitstable}).

%**********************************************************
%%Figure: Scatter
%
\begin{figure}%[b]
\centering
\includegraphics[width=0.45\textwidth]{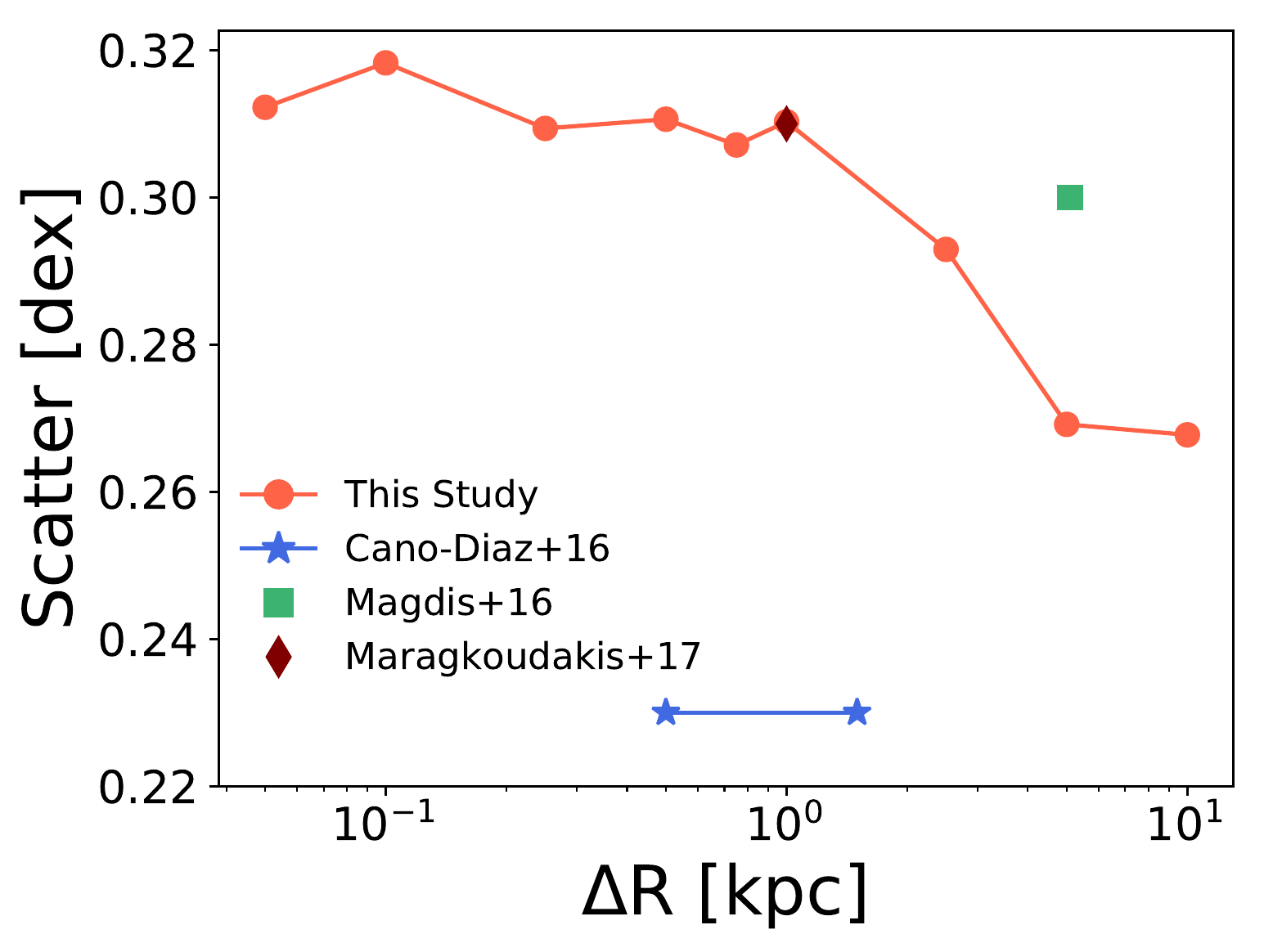}
\caption{Scatter of the local SFMS measured on various spatial scales (in kpc). Our own estimates are shown by orange dots.  SFMS scatter estimates by \cite{cano16}, \cite{magdis16}, and \cite{marag17} are shown in blue, green, and maroon, respectively.}
\label{fig:scatter}
\end{figure}
%
%**********************************************************

%-----------------------------------------------------SPATIAL RESOLUTION BINNING------------------------------------------------%
\section{Spatial Resolution of the SFMS}
\label{sec:reso}

The star formation law ($\Sigma_{\mathrm{SFR}}\propto \Sigma_{gas}^{N}$) has received considerable attention (\citealt{schmidt59,kennicutt98a,kravtsov03,schaye08}). 
However, only recently has it become possible to study scales down to 30~pc and address on what physical scales the star formation law holds and possibly even breaks down (\citealt{kravtsov03,onodera10,schruba10,kruijssen14,khop17}).

Measurements for the SFMS may no longer be appropriate on scales of individual giant molecular clouds (GMCs), particularly below 100~pc.
On such small scales, young clusters often drift from their parent GMC, thus disconnecting measurements of $\Sigma_{\mathrm{SFR}}$ from their related $\Sigma_{gas}$, and ultimately yielding incorrect \mstel\ and SFR.  
Furthermore, the study of individual molecular clouds lacks the sampling completeness of young stellar populations that is often assumed in SFR (and \mstel) transformations. SF activity can also vary from cloud to cloud, and since most SFR transformations are generalized to be applied across all environments, sampling of statistically robust and extended SF activity is therefore required.

\cite{kruijssen14} suggest that the smallest region size that will contain a complete sampling of SF environments is 490~pc.  Over such a scale, radiation from SF will propagate over the appropriate time scale for stellar mass growth.  They also derive the minimum scale that completely samples IMFs used in most transformation methods to be 340~pc.  The maximum spatial scale that young stellar clusters will drift from their parent GMC also appears to be 140~pc. The maximum of these spatial scales must therefore be used as the minimum scale over which to apply \mstel\ and SFR transformation, in order to produce physically meaningful measurements. Therefore we opt for a minimum spatial resolution of 500~pc in order to produce reliable measurements and probe local variations of \mstel\ and SFR within galaxies.

Finally, \cite{boquien15} examined SFR transformations on spatial scales from 33 to 2084~pc and posited that monochromatic SFR transformations break down before hybrid transformations with decreasing spatial scales. 
This further motivates our choice of the \Ca\ H$\alpha$+24$\mu$m transformation, whose calibration cover a large range of resolutions, over pure H$\alpha$, 23$\mu$m, or 24$\mu$m transformations.

\Fig{binning} shows our local SFMS binned on scales ranging from 0.05 to 10~kpc.  The slopes and zero-points of the different SFMS probed at different scales remain essentially the same.  The scatter of the SFMS at each resolution scale, shown in \Fig{scatter}, is also relatively constant out to 0.5-1~kpc, dropping slightly beyond that range.  This is broadly consistent with findings by \cite{kruijssen14} and we also adopt 500~pc as the nominal resolution scale for our study.

%----------------------------------------------------------RESULTS-----------------------------------------------------%
\section{The Star Formation Main Sequence}
\label{sec:results}

\subsection{The Global SFMS Relation}
\label{sec:global}

%**********************************************************
%%Figure: SFMS
%
\begin{figure}%[t!]
\centering
\subfigure[]{\includegraphics[width=0.45\textwidth]{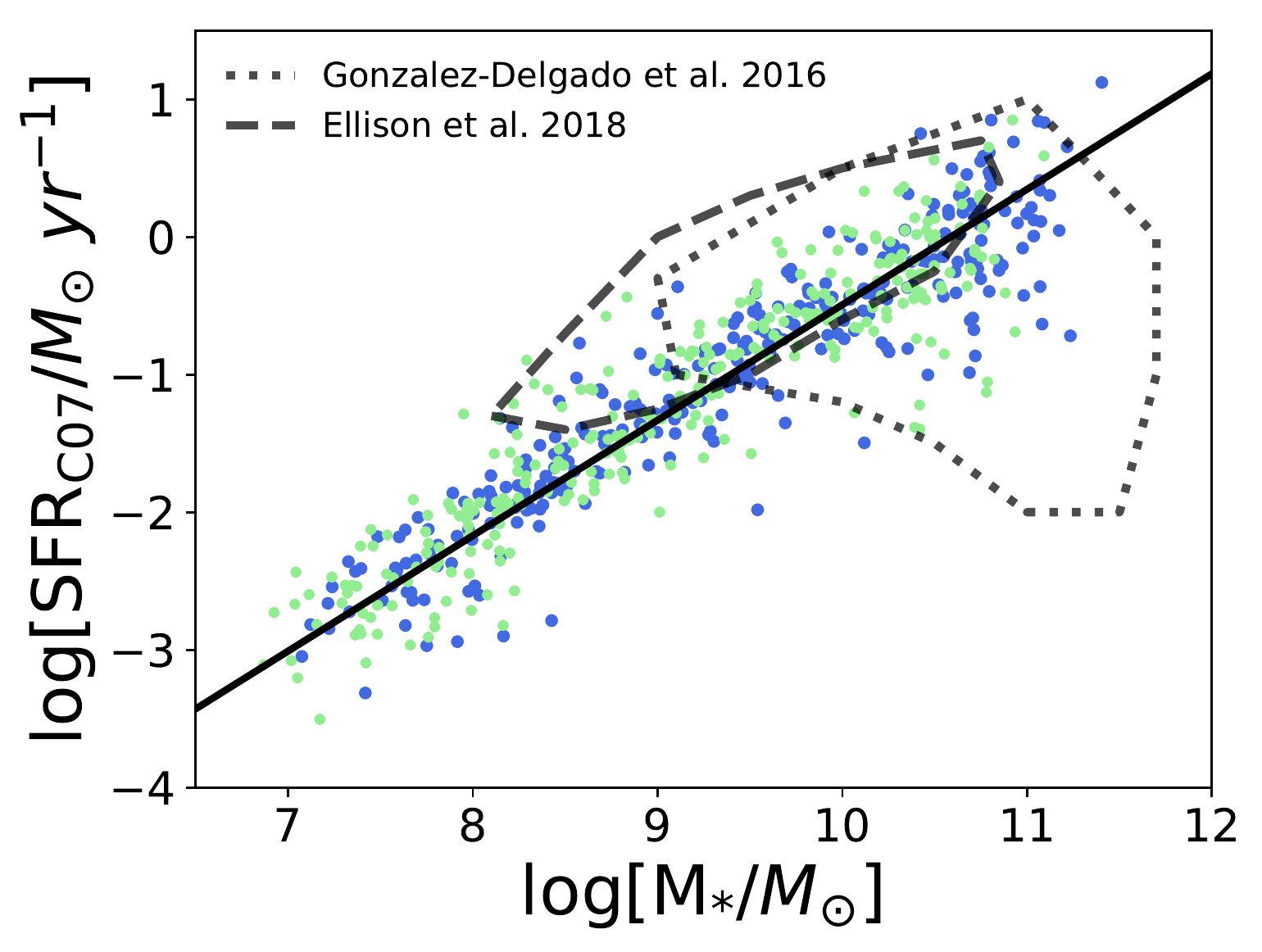}}
\hfill
\subfigure[]{\includegraphics[width=0.45\textwidth]{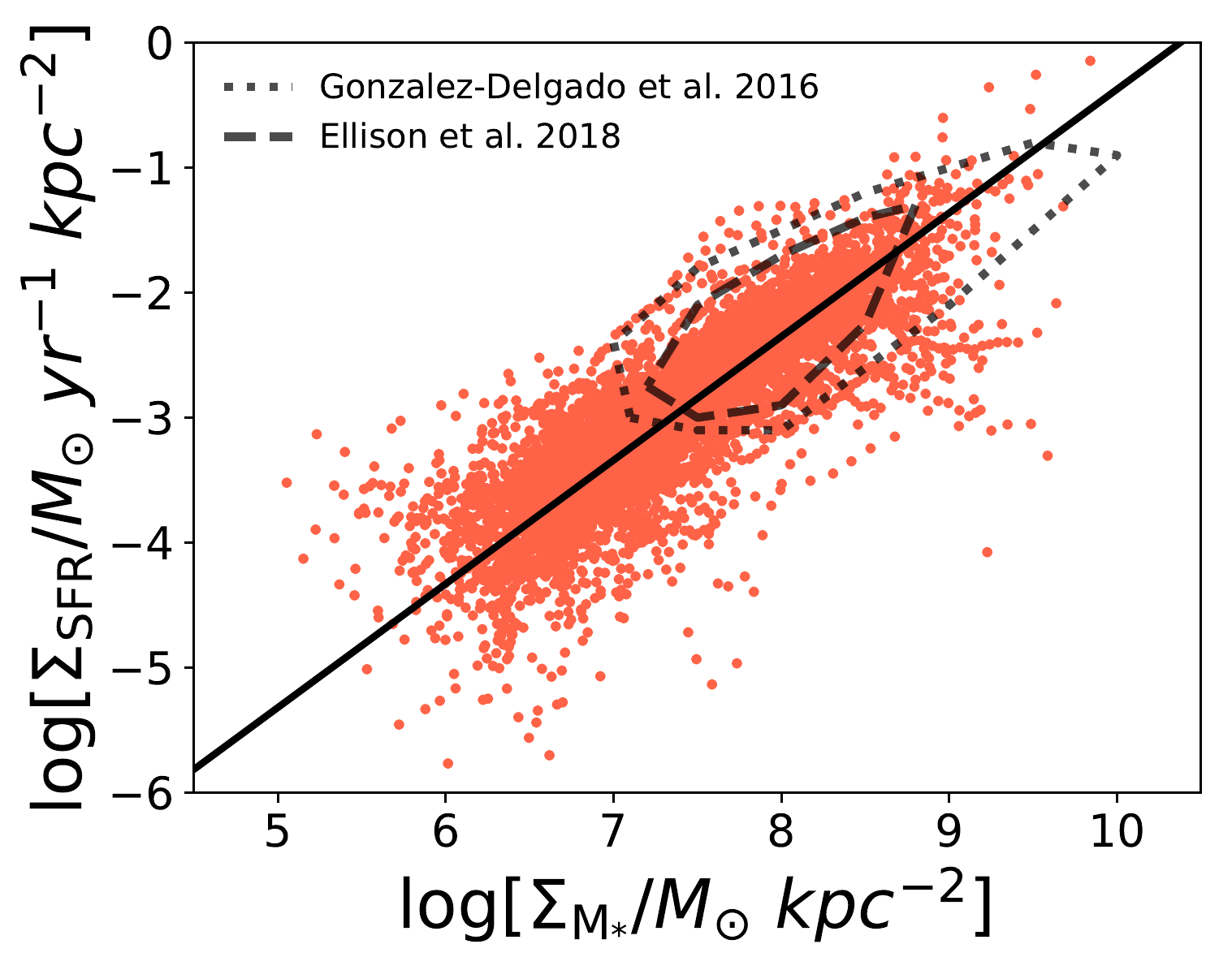}}
\caption{Global (top) and local (bottom) SFMS. \mstel\ values are calculated with a constant $M_{*}/L_{W1}=0.5$; the SFR are derived from the \cite{calzetti07} transformation. 
Local values for each galaxy are represented as a local density ($\Sigma_{\mstel}$, $\Sigma_{\mathrm{SFR}}$) and shown in orange, integrated values (\mstel, SFR) out to $R_{\mathrm{eff}}$ are in green, and total integrated galactic values are in blue. 
Solid lines are the linear fits to global and local values. Fits to the global (blue) and local (orange) SFMS are listed in the fourth row of \Table{fitstable}. 
Dashed lines represent the region within the SFMS for measurements found by \cite{gonzalez16} and \cite{ellison18}, as indicated in the legend.
}
\label{fig:theSFMS}
\end{figure}
%
%**********************************************************
%%Figure: All slopes
%
\begin{figure*}[t]
\centering
\includegraphics[width=1.0\textwidth]{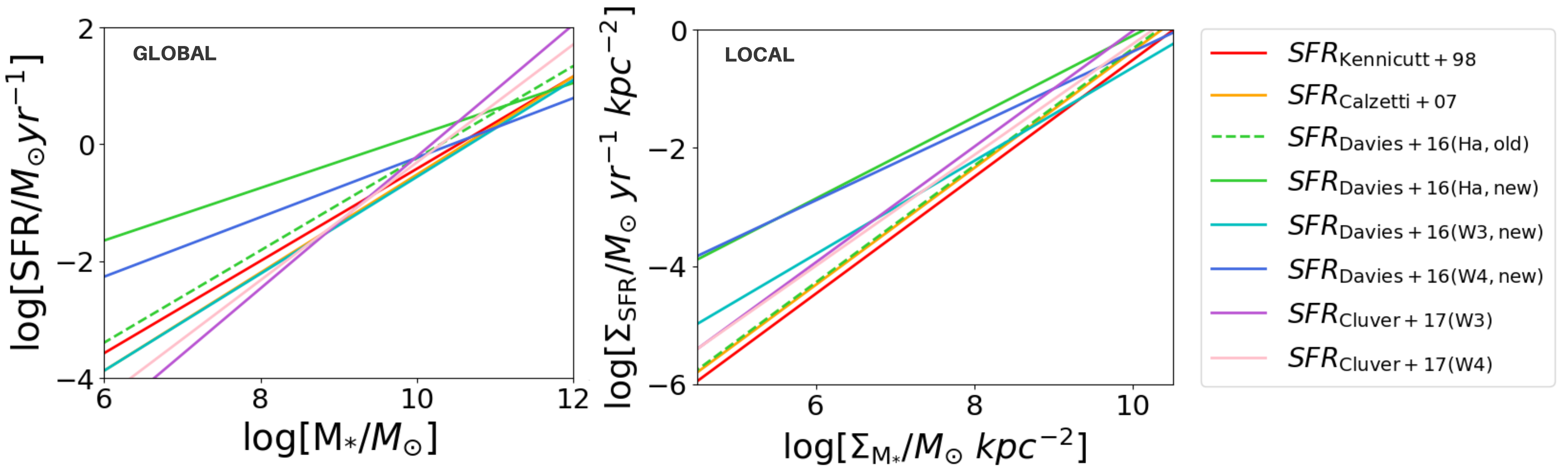}
\caption{SFMS fits to the global $\log{(\mathrm{SFR})} = a\log{(\mstel)} +b$ and local $\log{(\Sigma_{\mathrm{SFR}})} = a\log{(\Sigma_{\mstel})} +b$ relation, using various SFR transformation methods. \mstel values are calculated with a constant $M_{*}/L_{W1}=0.5$. Fits to the global SFMS are on the left panel, local SFMS fits on the right panel. Compare to \Fig{theSFMS}.}
\label{fig:allSlopes}
\end{figure*}
%
%**********************************************************
 %Table: Main Sequence Fits
%
\begin{table*}%[t!]
\centering
\caption{Star Formation Main Sequence Fit Parameters}
\begin{tabular}{ | c | c c c c | }
\hline
 & SFR Transformation & Slope $a$ & Zero Point $b$ & Standard Deviation $\sigma$ \\ [0.5ex]
\hline\hline
Global (G): & \citealt{kennicutt98b} 	 		& 0.79 	& -8.32  	& 0.31 \\
Local (L):  	& (Eq.~\ref{eq:KSFR})			& 0.99 	& -10.41 	& 0.37 \\
\hline
G:          	& \citealt{cluver17} \textit{W}3 		& 1.13 	& -11.51 	& 0.19 \\
L:          		& 	(Eq.~\ref{eq:ClSFR1})				& 0.98 	& -9.82 	& 0.26 \\
\hline
G:          	& \citealt{cluver17} \textit{W}4 		& 1.01 	& -10.41  	& 0.28 \\
L:          		& (Eq.~\ref{eq:ClSFR2})				& 0.94 	& -9.64 	& 0.35 \\
\hline
G:          	& \citealt{calzetti07} 		    			& 0.84 	& -8.92  	& 0.30 \\
L:          		& (Eq.~\ref{eq:CSFR}) 				& 0.99 	& -10.26 	& 0.30 \\
\hline
G:          	& \citealt{davies16}                    		& 0.79 	& -8.14  	& 0.31 \\
L:          		& 	H$\alpha$, old							& 0.99 	& -10.22 	& 0.37 \\
\hline
G:          	& \citealt{davies16}                    	& 0.45 	& -4.35  	& 0.21 \\
L:          		& 	H$\alpha$, new						& 0.69 	& -7.00  	& 0.30 \\
\hline
G:         	 	& \citealt{davies16}                  	& 0.83 	& -8.86  	& 0.16 \\
L:          		&  \textit{W}3, new					& 0.79 	& -8.45  	& 0.20 \\
\hline
G:          	& \citealt{davies16}                   	& 0.51 	& -5.33  	& 0.19 \\
L:          		& 	\textit{W}4, new					& 0.63 	& -6.67  	& 0.24 \\
\hline
\end{tabular}
\begin{tablenotes}
\item \textbf{Notes.} Numerical values for the global and local SFMS fits as plotted in \Fig{allSlopes}, using a constant $M_{*}/L_{W1}=0.5$ to calculate the stellar masses, and various transformation methods to determine SFRs. Fits are applied to the global $\log{(\mathrm{SFR})} = a\log{(\mstel)} +b$ and local $\log{(\Sigma_{\mathrm{SFR}})} = a\log{(\Sigma_{\mstel})} +b$ relations. The global and local SFMSs for the \cite{calzetti07} SFR transformation (fourth row) are displayed in \Fig{theSFMS}.
\end{tablenotes}
\label{fitstable}
\end{table*}
%
%**********************************************************

The global SFMS is calculated using total integrated values for each galaxy, which are determined from total enclosed SINGG and \textit{WISE} measurements integrated out to the same, matched radius. Multiple transformations were examined, as outlined in \sec{methods}.  \Fig{theSFMS} displays the SFMS constructed with stellar masses using a constant $M_{*}/L_{W1}=0.5$ and SFRs from the \Ca\ transformations (\eq{CSFR}). Refer back to \sec{sfrs} \& \sec{stellm} for the motivation for these transformations.
The \Esk\ and \Cla\ \mstel\ transformations yield essentially the same results (see additional SFMS results in App.~C).

The top panel of \Fig{theSFMS} shows the SFMS integrated values are overlayed in blue and green.  Local values, plotted in orange in the bottom panel, will be discussed in \sec{local}. The blue points represent the total integrated measurement out to the maximum radius of detection, while green dots are measurements integrated out to $R_{\mathrm{eff}}$. The blue and green sequences are roughly the same ($a_{\mathrm{blue}}=0.84$ vs. $a_{\mathrm{green}}=0.84$, $b_{\mathrm{blue}}=-8.92$ vs. $b_{\mathrm{green}}=-8.81$, and  $\sigma_{\mathrm{blue}}=0.30$ vs. $\sigma_{\mathrm{green}}=0.29$), demonstrating that regions beyond $R_{\mathrm{eff}}$ do not considerably affect the global SFMS, and the choice of maximum radius has minimal impact on the global SFMS. In \sec{local} \& \ref{sec:scatter} we will compare the SFMS in inner and outer regions.  The strong positive linear correlation in the global SFMS already reported by others (e.g. \citealt{noeske07,speagle14,salmon15}) is reproduced here with SINGG and \textit{WISE} data. Linear fits to the SFMS with the equation $\log{(\mathrm{SFR})} = a\log{(\mstel)} +b$ using different SFR and  \mstel\ transformations are listed in \Table{fitstable}, and the global fit is overlayed in solid black in \Fig{theSFMS}.

Our measured slopes ($a$), ranging from about 0.8 to 1.1, are consistent with those reported by \cite{noeske07,wuyts11,speagle14,tomczak16}, accounting for evolutionary factors, excluding fits with the re-calibrated \ha\ and \textit{W}4 SFR transformations by \cite{davies16}. The wider range of slopes in \cite{speagle14} is largely due to the broad range of environments encompassed, including starbursting to quenched environments. Similarly, the trend of decreasing SFRs with lower redshift \citep{noeske07,tomczak16} results in our magnitude of SFRs, consistent with \cite{wuyts11} at $z\approx0.02$-$0.2$.
However, \Table{fitstable} makes clear that our fits, even for the same data set, are not constant. This suggests that the SFMS error budget may be largely spoiled by systematic errors in the adopted transformations, rather than intrinsic environmental factors. The regimes of global SFMS measurements by \cite{gonzalez16} and \cite{ellison18} are outlined in the top panel of \Fig{theSFMS} in gray, and lie approximately within the same regime as our data, whereas the CALIFA and MaNGA samples are restricted to $\log{(\mstel)}>9.0$ systems \citep{walcher14,ellison18}. The CALIFA sample also has a larger contribution of early-type, quiescent galaxies, which accounts for the lower SFR measurements at high \mstel\ by \cite{gonzalez16}. While our SFRs are somewhat smaller than those of \cite{ellison18}, they are broadly consistent with \cite{gonzalez16} and all agree within the level of uncertainty attached to these transformations (about a factor of two). The higher range of stellar masses show broad consistency with all other considered samples.

The left panel of \Fig{allSlopes} shows the varying global SFMS fits (slope and intercept) produced using different SFR transformations, as discussed in \sec{sfrs}.  Four of them were taken from \cite{davies16}. With the exception of the solid green and solid blue lines from \cite{davies16} for H$\alpha$ and \textit{W}4 transformations, the various SFMS fits appear to be highly consistent with each other.  
However, the calibrations by \cite{davies16}, originally calibrated for stellar masses above $10^{9}$~M$_{\odot}$, produce unreasonably high SFRs at lower stellar masses. 
This figure otherwise suggests that the common suite of transformations compares favorably on global, integrated scales.
Interestingly, the \Clb\ \textit{W}3 and \textit{W}4 transformations have higher SFRs compared to other transformations at high \mstel\ values. This could arise from the higher dust content at high SFR and stellar mass, and the \textit{W}3 and \textit{W}4 may perform better in this regime. The \textit{W}3 and \textit{W}4 transformations could therefore be more robust on global scales, while a hybrid calibration would be more suitable for local measurements.

\subsection{The Local SFMS Relation}
\label{sec:local}
%**********************************************************
%%Figure: HI mass residual
%
\begin{figure}%[t]
\centering
\includegraphics[width=0.42\textwidth]{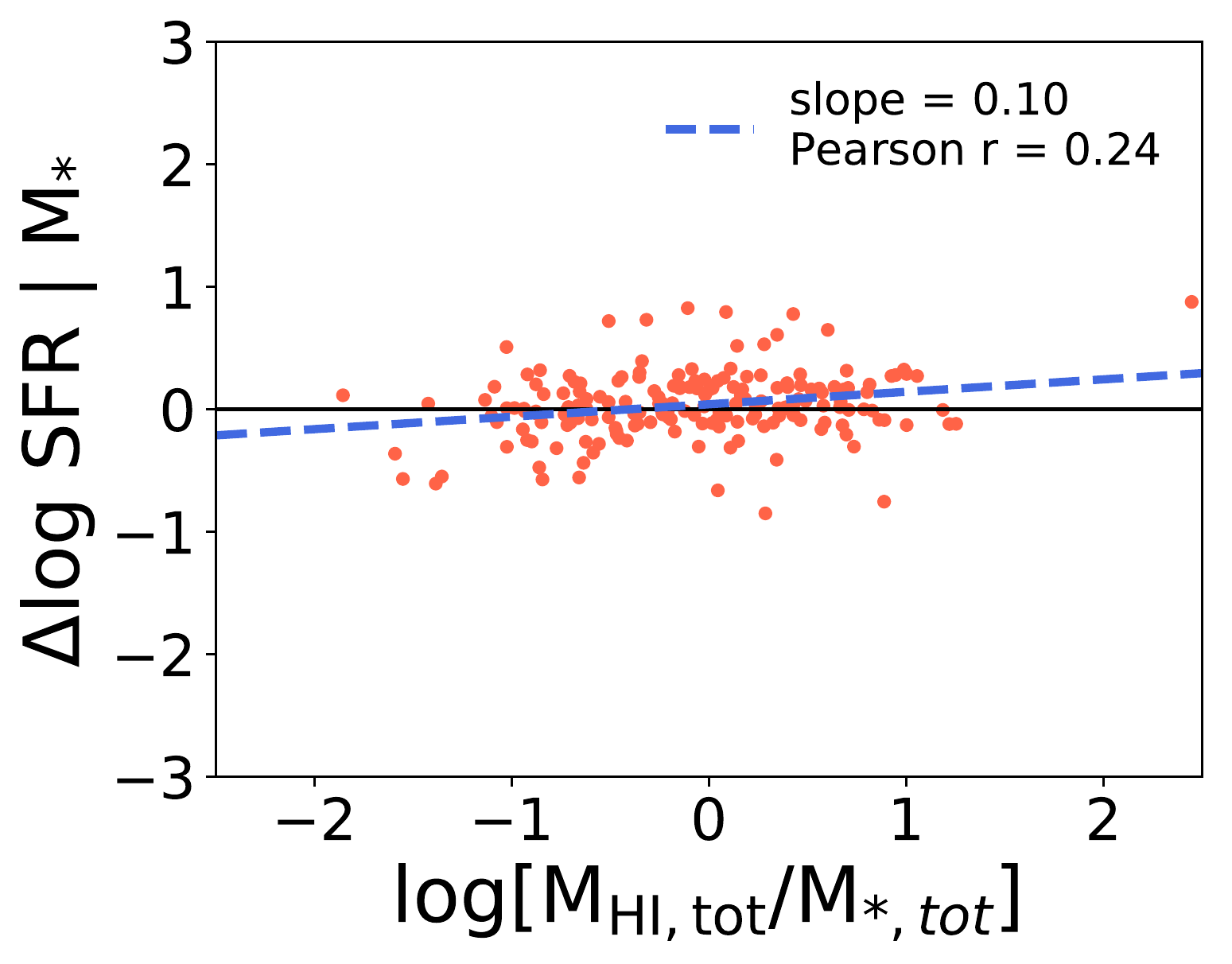}
\caption{Vertical distance from global SFMS fit ($\Delta$log(SFR)) at a fixed \mstel\ against the total \hi\ mass fraction (M$_{\hi,tot}$/M$_{*,tot}$). The black line is the linear fit to the SFMS; the blue line is the fit the residuals corresponding to each global parameter. The correlation coefficient is given by Pearson r.}
\label{fig:mhi}
\end{figure}
%
%**********************************************************
%%Figure: T-type + Scatter
%
\begin{figure*}[t!]
%\captionsetup{width=0.9\textwidth}
\centering
\subfigure[]{\includegraphics[width=0.45\textwidth]{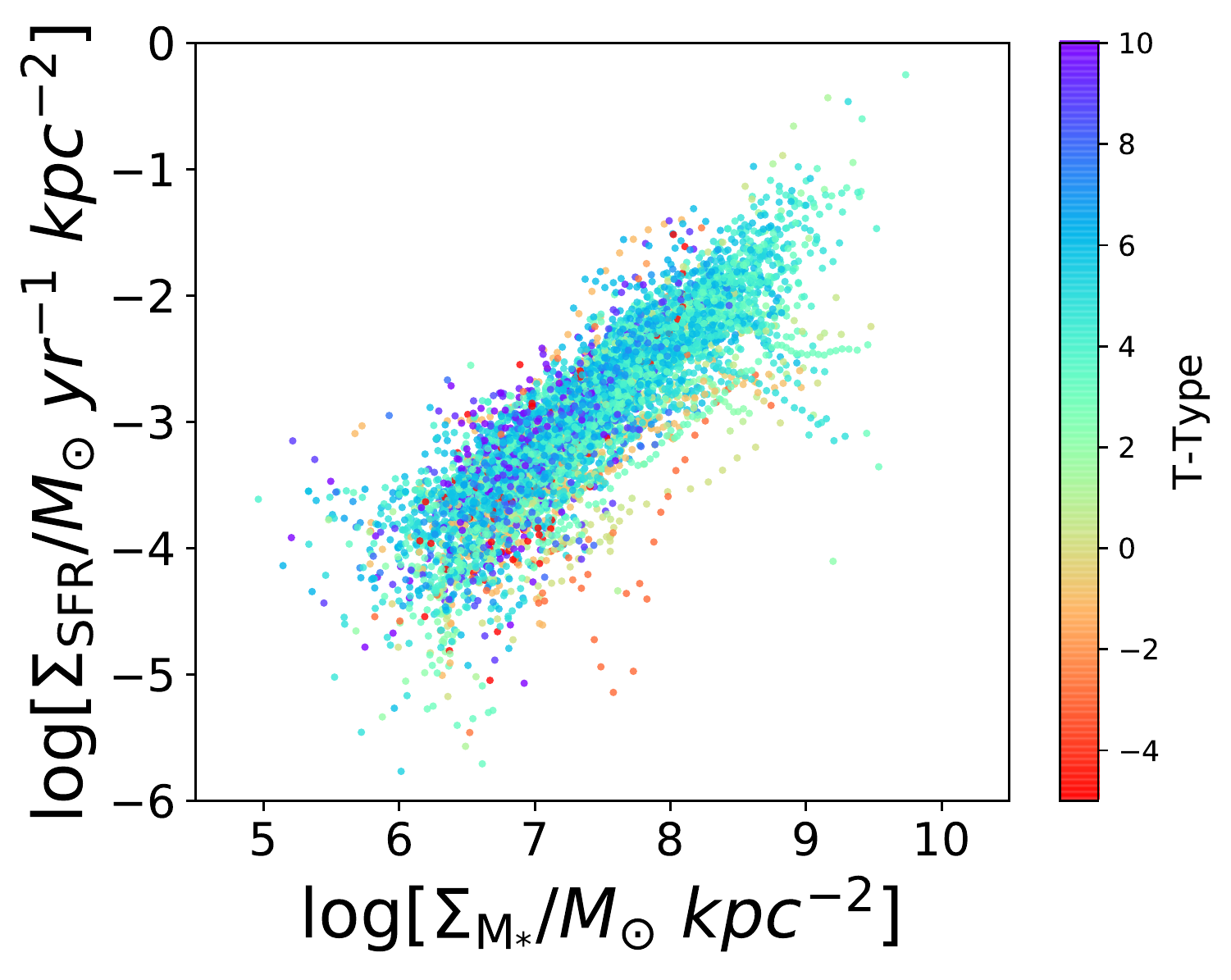}\label{fig:morphgrad}}
\hfill
\subfigure[]{\includegraphics[width=0.45\textwidth]{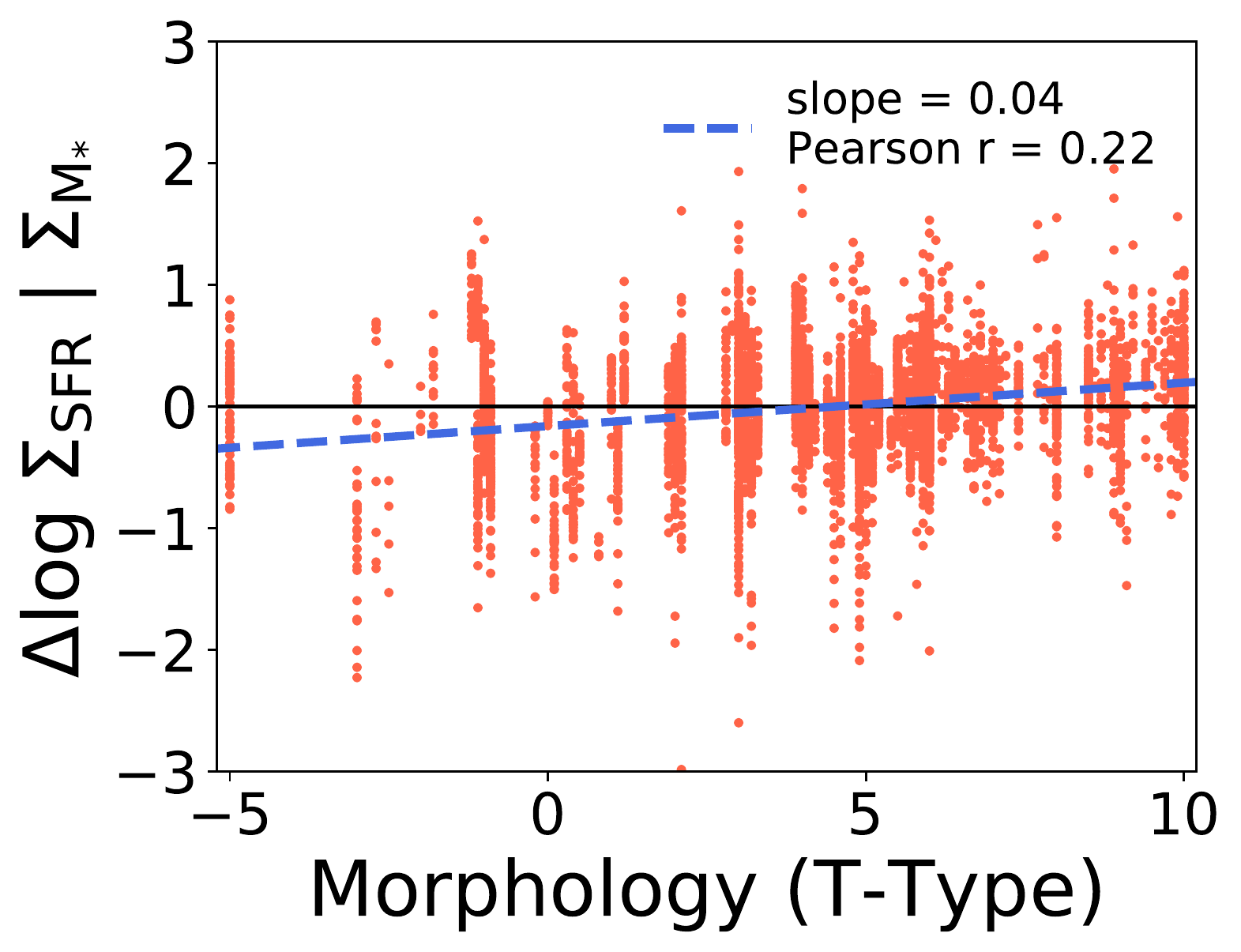}\label{fig:morphres}}
\caption{\textbf{Left:} Distribution of galaxy morphology (T-Type) across the local SFMS. 
\textbf{Right:} Vertical distance from local SFMS fit ($\Delta\log{(\Sigma_{\mathrm{SFR}})}$) at a fixed $\Sigma_{\mstel}$ scaled against Morphology type (T-Type). The black line is the linear fit to the SFMS; the blue line is the fit of the residuals corresponding to scaled radius. The correlation coefficient is given by Pearson r.}
\label{fig:morph}
\end{figure*}
%
%**********************************************************

The local SFMS, seen in the bottom panel of \Fig{theSFMS} in orange, was derived from radially resolved profiles of $\Sigma_{\mstel}$ and $\Sigma_{\mathrm{SFR}}$ for each galaxy, using transformations by \Ca\ and \Cla; the motivation for these is presented in \sec{methods} \& \ref{sec:global}). Each measurement represents an annular ring centered around the galactic center, and each ring has a width of $\approx500$ pc by design.
The strong positive, linear correlation between SFR and \mstel\ typically seen in the the global SFMS is also evident on local $\Sigma_{\mstel}$ and $\Sigma_{\mathrm{SFR}}$ scales. The local fit in the bottom panel of \Fig{theSFMS} is shown by black solid line. This further confirms that these two tracers of star formation are strongly coupled not only amongst galaxies but within themselves; their variation closely track each other. The SFMS regime studied by \cite{gonzalez16} and \cite{ellison18} is again outlined in the bottom panel of \Fig{theSFMS} in gray, and overlaps with our data. However, unlike the CALIFA and MaNGA samples, significantly lower values of $\Sigma_{\mstel}$ and $\Sigma_{\mathrm{SFR}}$ measurements are probed in our study thanks to our inclusion of lower stellar mass systems.

Our local SFMS slopes average $a\sim1$ and are slightly steeper than integrated SFMS slopes (\Table{fitstable}). They are also consistent, though at the high end, with reported slopes ($a\sim0.7-1.0$) from other studies of the spatially resolved SFMS (\citealt{perez13,wuyts13,hemmati14,cano16,gonzalez16,magdis16,abdurro17,marag17}).
The right panel of \Fig{allSlopes} shows that SFMS fits from a variety of SFR transformations are highly consistent for local measurements, the exception being the green, cyan, and blue solid lines corresponding to the \cite{davies16} relations. This can likely be attributed to the fact that these new calibrations were applied to higher mass systems and were not spatially resolved, whereas the \Ken\ and \Ca\ transformations have been calibrated to such resolved environments. GAMA \ha\ measurements in \cite{davies16} also have a limited aperture size of 2 arcsec that may cause systematic offsets, and the \textit{W}4 detections were quite poor due to the redshift limits ($z\lesssim0.13$).
Although environment becomes more stochastic at lower \mstel\ regimes, it is interesting that these local SFMS fits and their scatter remain consistent. There is a slight increase in scatter about spatially resolved measurements and \sec{scatter} investigates the influences by regions within a galaxy versus at a global scale more closely.

%----------------------------------------------------------SCATTER-----------------------------------------------------%
\section{SFMS Scatter Analysis}
\label{sec:scatter}

The main focus of our investigation is to gain a better understanding of the intrinsic scatter of the local, spatially resolved SFMS, since SF shows a strong local scale dependence by the star formation law. The observed scatter about the spatially resolved SFMS is comparable to, though slightly larger than, the scatter about the global SFMS with integrated \mstel\ and SFR values, as evidenced from \Fig{theSFMS} and \Table{fitstable}. 
Galaxies in our study are star-forming, by virtue of the SINGG survey selecting galaxies with \hi. While the presence of \hi\ does not directly imply SF, it often correlates with the presence of \ha\ at a global scale, which is representative of SF, and this is evident in our sample.
Though, we have not imposed that all regions within each galaxy be star-forming. While this may contribute to the scatter of the local SFMS, the division between distinct star-forming and quiescent regions that is seen in some global studies of the SFMS  (\citealt{noeske07,eales17,pandya17}) does not exist here, as mentioned in \sec{global}. Thus, while SF activity may fluctuate across a galaxy, within a star-forming galaxy the extremes are not sufficiently large to form two distinct SF sequences.

\cite{cano16} eliminated regions that do not satisfy the ``star forming conditions'' (set by equivalent H$\alpha$ width measurements, a common criteria for global scale SFMS studies) from their spatially resolved analysis.  As such quiescent regions were not weeded out from our analysis, this may explain the larger (perhaps more physical) scatter that we measure.

Our approach to analyzing the scatter includes the graphical representation of the SFMS as a function of several fundamental parameters and the production of residual figures, emulating the work of \cite{dutton07} and \cite{brennan17}.  Key parameters for our scatter analysis include: galactic location ($R/R_{\mathrm{eff}}$), morphology (T-Type), bulge-to-disk ratio, concentration index, effective surface brightness (SB$_{\mathrm{eff},W1}$), inclination, $\Sigma_{\mathrm{SFR}}$, $\Sigma_{\mstel}$, and so on. Most of these parameters display no strong correlation with the SFMS scatter (see App.~D for corresponding plots). However, interesting trends with morphological type and radius emerge for individual galaxies.

\subsection{\hi\ Mass Dependence}
\label{sec:hi}

We first examine any correlation of the SFMS with the total \hi\ mass. 
Our galaxy sample is drawn from the SINGG, which itself targets were \hi -rich galaxies selected in HIPASS.
Therefore, we wish to assess if this selection biases our sample in any relevant way. Of particular interest is the claim by \cite{saintonge16} of a correlation between the ratio of total \hi\ mass to total stellar mass of a galaxy and its position (scatter) within the global SFMS.
However, our own examination of the SFMS residuals with \hi\ mass (\Fig{mhi}) reveals no such dependence for our sample. This suggests, at least for our sample of star-forming galaxies (with few, if any, quiescent early-type galaxies), that the role of the \hi\ mass in regulating star formation is minimal.

\subsection{Morphological Dependence}
\label{sec:morph}
%**********************************************************
%%Figure: Radius + Scatter
%
\begin{figure*}[t!]
%\captionsetup{width=0.9\textwidth}
\centering
\subfigure[]{\includegraphics[width=0.45\textwidth]{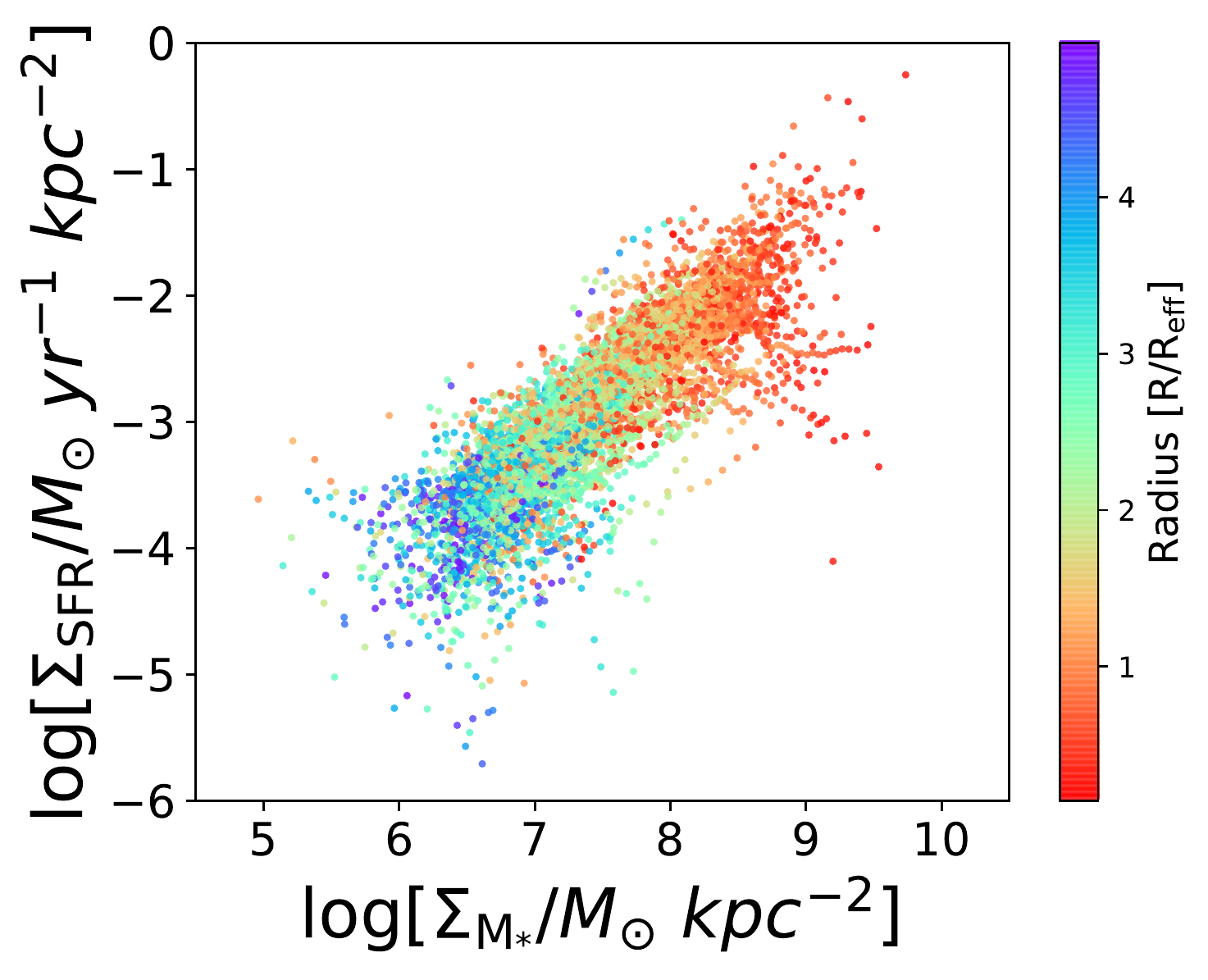}\label{fig:radialscatter}}
\hfill
\subfigure[]{\includegraphics[width=0.45\textwidth]{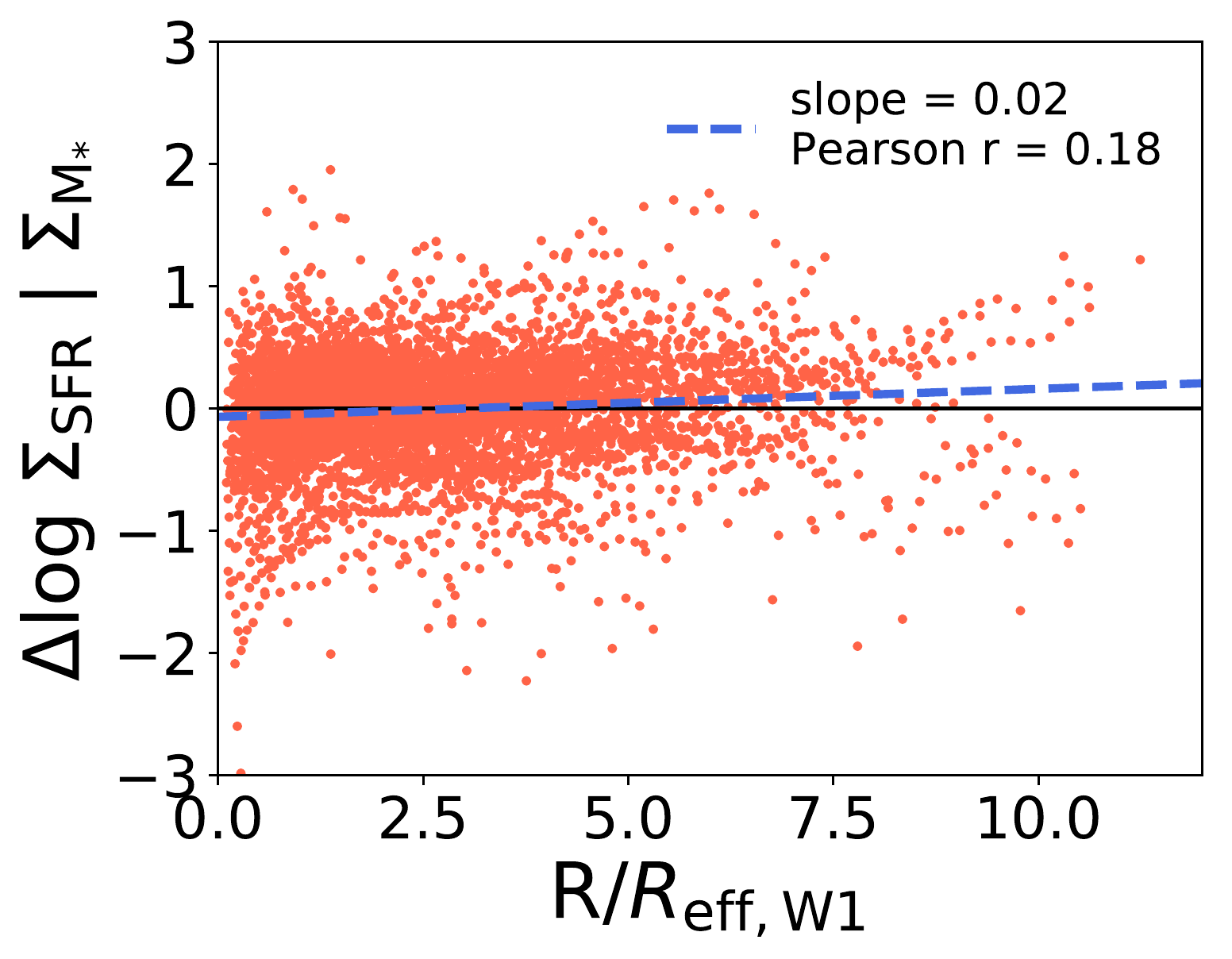}\label{fig:radRes}}
\caption{\textbf{Left:} Distribution of radial location ($R/R_{\mathrm{eff}}$) within the local SFMS. The gradient displays the range of radial values from $R/R_{\mathrm{eff}}=0$ to $R/R_{\mathrm{eff}}=5$.
\textbf{Right:} Vertical distance from local SFMS fit ($\Delta\log{(\Sigma_{\mathrm{SFR}})}$) at a fixed $\Sigma_{\mstel}$ scaled against radial value ($R/R_{\mathrm{eff}}$). The black line is the linear fit to the SFMS; the blue line is the fit of the residuals corresponding to scaled radius. The correlation coefficient is given by Pearson r.}
\label{fig:radius}
\end{figure*}
%
%**********************************************************

It is also necessary to examine any correlation with morphology, as CALIFA SFMS results continue to find strong dependence of the local SFMS scatter on morphology \citep{gonzalez16,gonzalez17,lopez18}.  According to them, morphology tracks perpendicularly to the SFMS on both global and local scales, suggesting that the host galaxy determines the general SFMS trends (slope and zero point) throughout all regions within the galaxy.  
While our morphological type coverage is not as extensive, by virtue of the HIPASS sample selection, we do have a broad selection of early-type to late-type to irregular galaxies. Our scatter analysis in fact yields a different trend. 
\Fig{morphgrad} shows that late-type galaxies (T-Types from $0$ to $7$) lie throughout the local SFMS, whereas very early-type (T-Type $-6$ to $-3$) and irregular (T-Type $8$ to $10$) galaxies appear to dominate in the low $\Sigma_{\mstel}$ regime. 
There is no clear orthogonal trend to this relation.  This contradictory trend that we find could result from early-type and irregular galaxies still showing a significant amount of observable SF (at least traced by the presence of HI by HIPASS), but not having enough mass build-up to contribute to the higher mass regions of the SFMS; and likely much smaller systems than typical early-type galaxies.   Sample selection is clearly key in establishing the trends observed in our respective analyses of morphological dependence. 

\subsection{Radial Dependence}
\label{sec:radscat}
%**********************************************************
%%Figure: Galaxy behavior Schematic 3
%
\begin{figure*}%[b]
\centering
\includegraphics[width=1\textwidth]{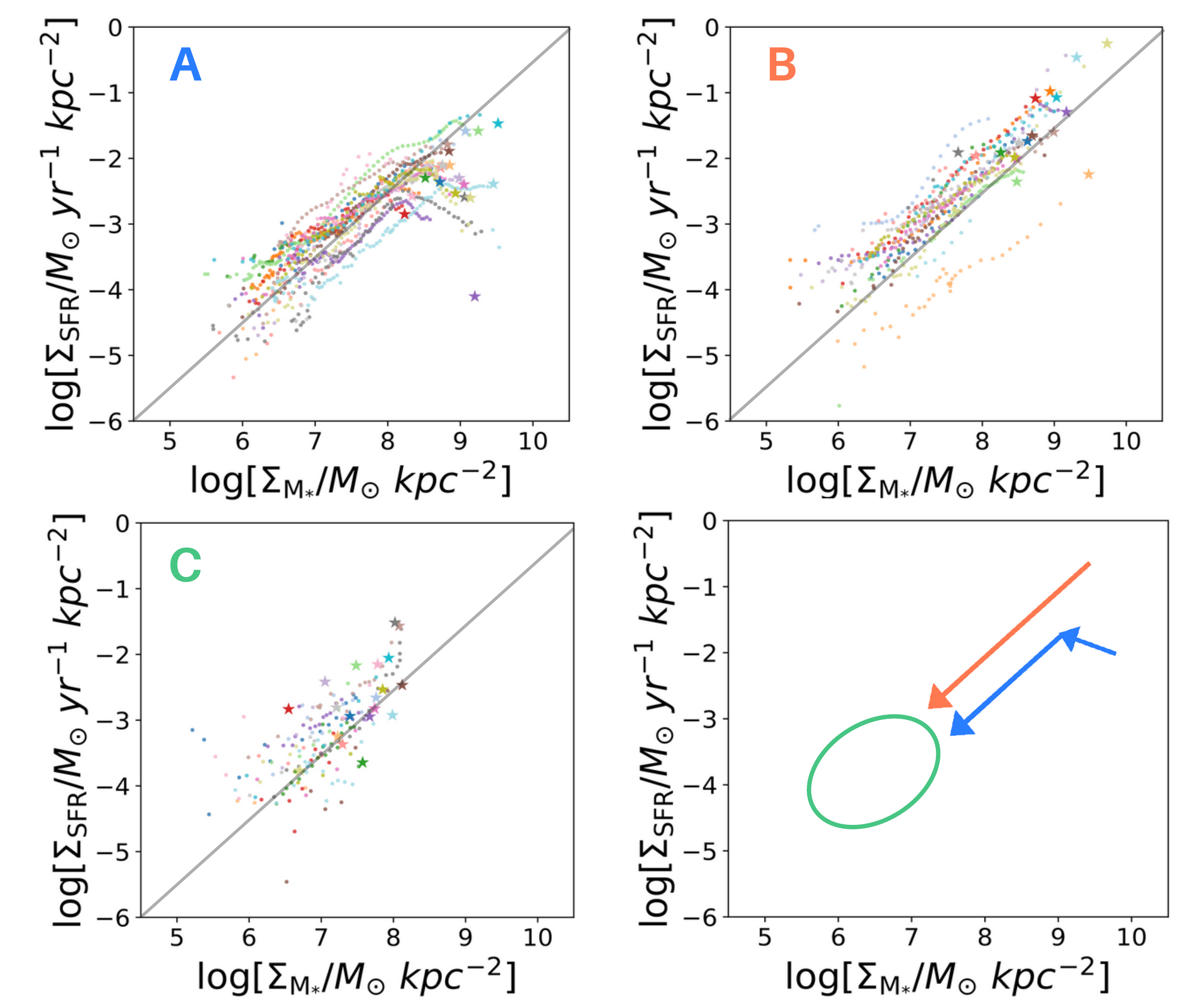}
\caption{Tracks of individual galaxies within the local SFMS, by distinct colors. The star point type indicates the central measurement, and the dots represent the radial values leading out from the center. Only 20 galaxies are plotted in each figure for clarity. The top two and bottom left panels highlight the general characteristic behavior of chosen galaxies across the SFMS; the bottom right panel presents an idealization of these trends. We have categorized galaxies into three distinct SFMS tracks: Type A (blue); Type B (orange); Type C (green), as depicted in the bottom right panel. Distinct features are discussed in the paper.}
\label{fig:tracks}
\end{figure*}
%
%**********************************************************
%%Figure: Galaxy behavior Schematic 3
%
\begin{figure*}%[b]
\centering
\includegraphics[width=1\textwidth]{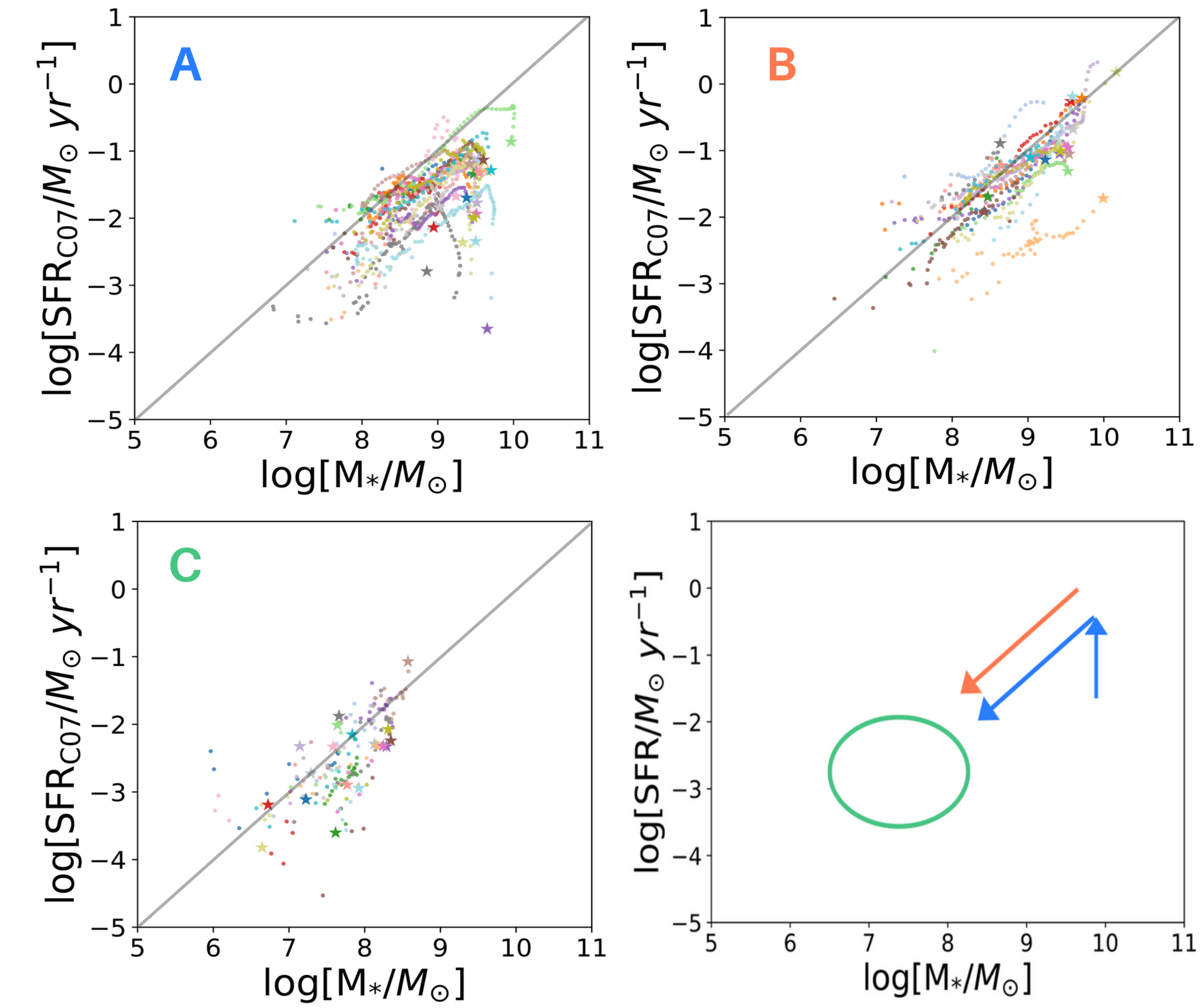}
\caption{Tracks of individual galaxies within the local SFMS, emulating \Fig{tracks}, but here showing absolute measurements of SFR and \mstel . This representation emphasizes the distinction between the radial behaviors (as discussed in text) of the types A and B, whereby the rise in SFR for the centers of type A systems occurs at constant stellar mass.
}
\label{fig:tracks2}
\end{figure*}
%
%**********************************************************

The radial dependence of the SFMS requires additional scrutiny. 
Gradient and residual trends are shown in \Fig{radialscatter} and \ref{fig:radRes} against the radial parameter, $R/R_{\mathrm{eff}}$. Little correlation is detected in these plots, as is true for most other SFMS residual analyses (see again App.~D).

However, patterns emerge when the individual paths of a random selection of galaxies, each distinguished by a unique color, are plotted in \Fig{tracks}. These reveal the unique behavior of each galaxy within the SFMS, where a galaxy's central data point is denoted by a star. Galaxies with local mass densities higher than log($\Sigma_{\mstel}$)$\sim$7.5 seem to define two trajectories as their tracks progress from the galactic center to their outskirts. One class (A) initially plateaus in $\Sigma_{\mathrm{SFR}}$ with decreasing $\Sigma_{\mstel}$.  Its radial track then decreases in both $\Sigma_{\mathrm{SFR}}$ and $\Sigma_{\mstel}$ in lock-step with the SFMS. The alternate class (B) bypasses the initial plateau in $\Sigma_{\mathrm{SFR}}$, plunging from high values of $\Sigma_{\mathrm{SFR}}$ and $\Sigma_{\mstel}$ in lock-step with the SFMS from the center out. Below log($\Sigma_{\mstel}$)$\sim$7.5, individual galaxy tracks lack structure and show random paths (C).  The scale of log($\Sigma_{\mstel}$)$\sim$7.5 appears to mark an important transition; whether a galaxy follows the tracks identified above or not, all trends vanish for local SFMS tracks below log($\Sigma_{\mstel}$)$\sim$7.5. This suggests distinctly different mechanisms for high and low mass density environments.  These three galactic behaviors are projected for clarity in the bottom right panel of \Fig{tracks}; we label them as the type A, B, and C galaxy groups. The schematics given in the bottom right panel of \Fig{tracks} are representative of the broad $\Sigma_{\mstel}$ and $\Sigma_{\mathrm{SFR}}$ regimes found in our sample; exceptions also exist. 

While density measurements eliminate the variation in area amongst all measurements, so as to better compare different galaxy sizes and internal/radial locations, absolute measurements lack the additional uncertainty introduced from distance errors with density values. We therefore present in \Fig{tracks2} a complementary representation of \Fig{tracks} to further emphasize the distinction between these three types. In this space, the type A population displays an initial sharp rise in the SFR (rather than plateau), whereas type Bs follow the SFMS at all radial points. The stochasticity in type C measurements is also clear in \Fig{tracks2}, the log($\Sigma_{\mstel}$)$\sim$7.5 in density space transitions to log(\mstel)$\sim$8.5 for absolute measurements.

The type A curved tracks provide compelling motivation for inside-out quenching in higher mass systems, as the $\Sigma_{\mathrm{SFR}}$ at the center appears to be significantly reduced compared to the corresponding high stellar mass density.  However, we now seek distinguishing features between this and the type B straight, seemingly unquenched, galaxy group that lies in the same local stellar mass density regime.  The type B galaxy group generally has higher $\Sigma_{\mathrm{SFR}}$ at a given $\Sigma_{\mstel}$, as evident in \Fig{tracks}.

Another striking feature that emerges from \Fig{tracks} is that the scatter for structures above log($\Sigma_{\mstel}$)$\sim$7.5 appears to be caused by a systematic offset between galaxies. Disregarding the initial rise in $\Sigma_{\mathrm{SFR}}$ among the type A galaxies, each galaxy appears to lie roughly parallel the SFMS fit, suggesting that the scatter is driven on global scales across galaxies. In fact, applying a linear fit to each individual galaxy track yields consistent slopes ($a$) with varying intercepts, and the scatter about the SFMS fit associated with an individual galaxy is less than the total scatter of the SFMS, supporting the notion that scatter is dominated by vertical offsets between galaxies, for type A and B galaxies.

By contrast, below log($\Sigma_{\mstel}$)$\sim$7.5, SFMS measurements appear to be stochastic in all parts of the galaxies. Interestingly, the overall scatter ($\sim$0.3 dex) remains constant at all $\Sigma_{\mstel}$.  It is likely that the random scatter in lower mass density systems is due to stochastic sampling of SF behaviors and environments at such small $\Sigma_{\mstel}$ scales, and the vertical offsets in type A and B galaxies would be due to varying global accretion rates. The age and accretion histories are difficult to quantify observationally, but stellar mass and SFR density distributions across a galaxy may potentially inform us of their impact. In this context, a future investigation of star formation and accretion histories with semi-analytic models is certainly warranted.
 
%**********************************************************
%%Figure: Histogram of pops by Mtot and Mhi
%
\begin{figure}%[t!]
\centering
\subfigure[]{\includegraphics[width=0.42\textwidth]{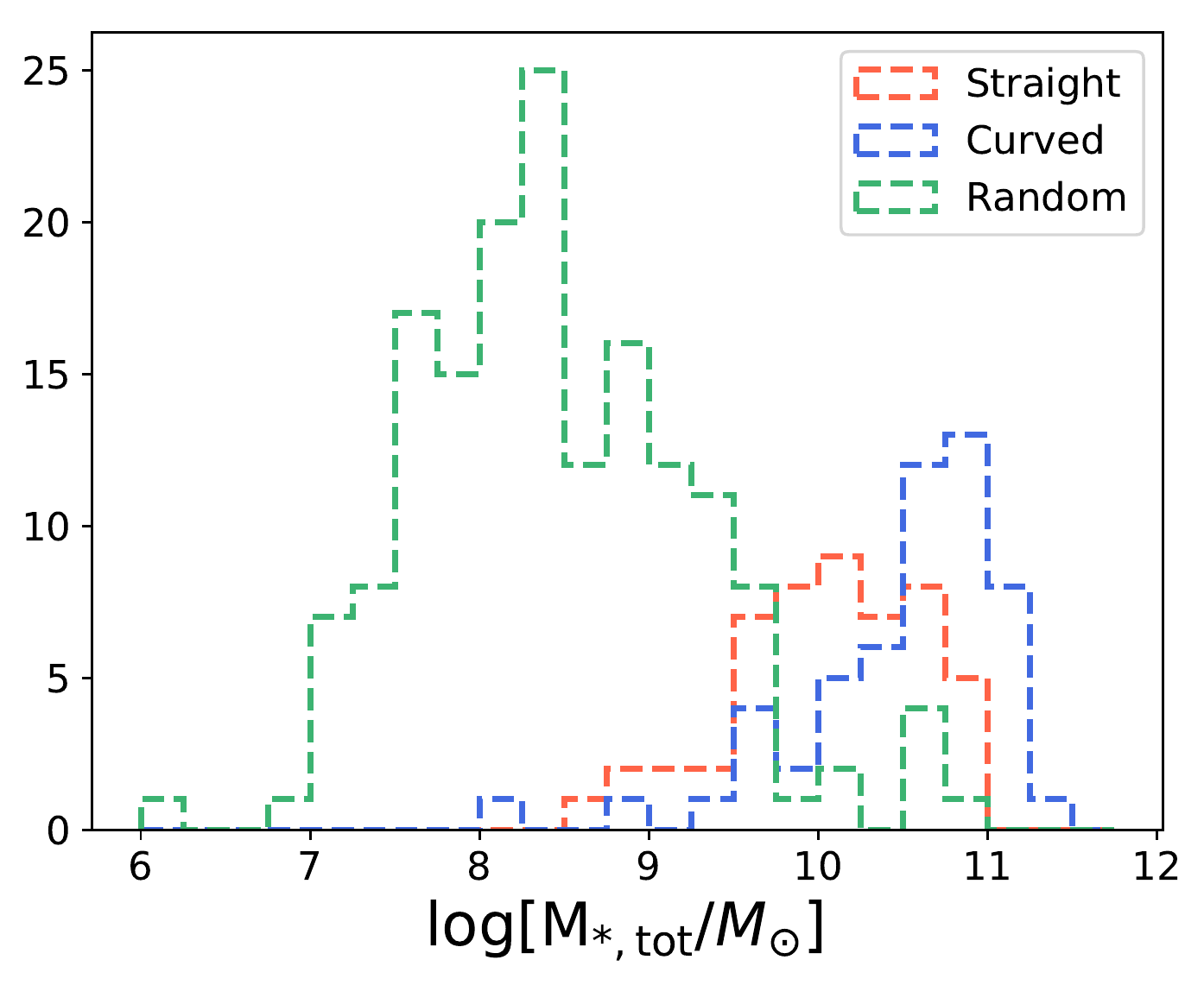}}
\vfill
\subfigure[]{\includegraphics[width=0.42\textwidth]{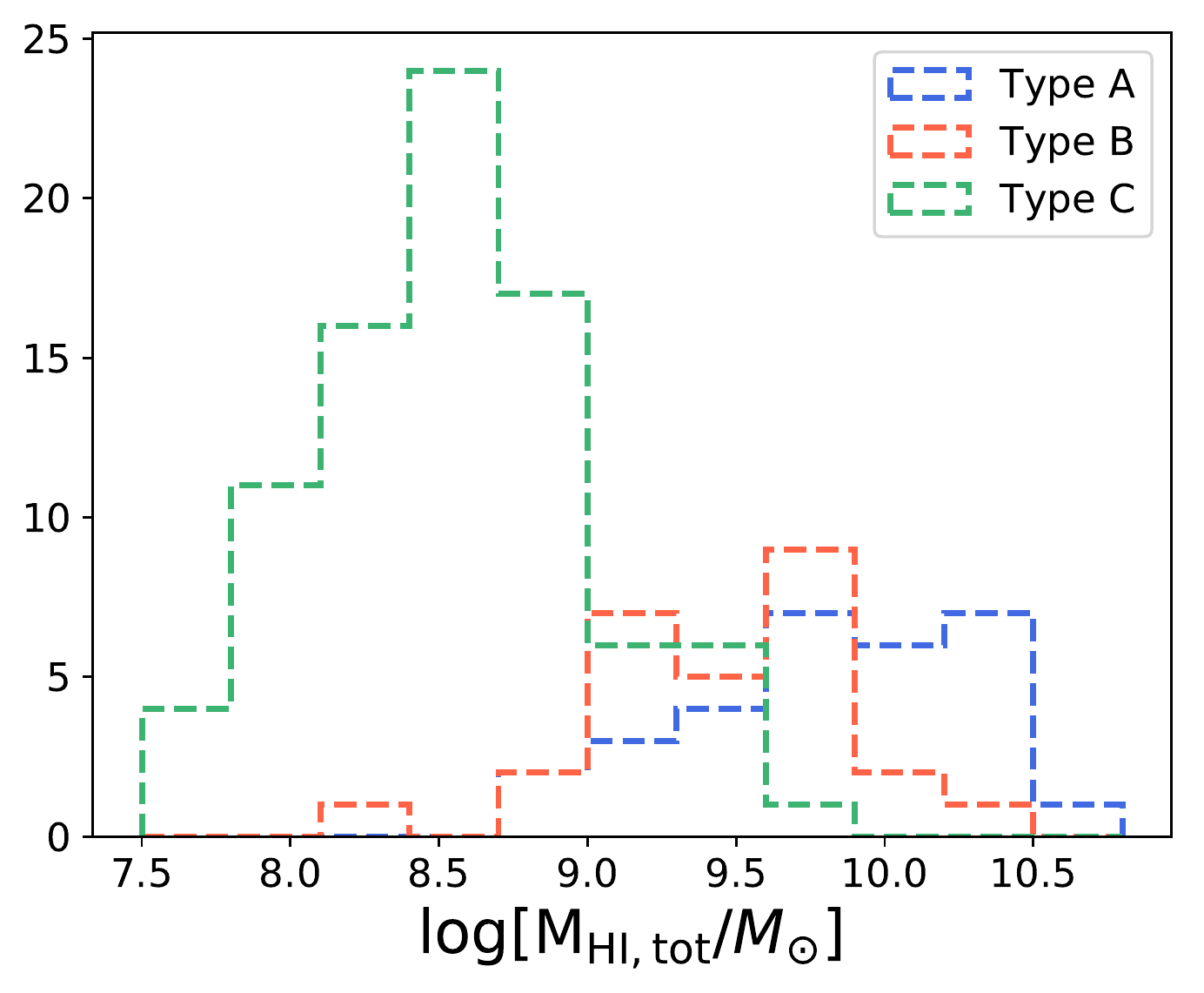}}
\caption{The distribution of (\textit{top}) total stellar mass (M$_{\mathrm{tot}}$) and (\textit{bottom}) total \hi\ mass (M$_{\mathrm{\hi}}$) of the three galaxy groups defined in \Fig{tracks}.}
\label{fig:Mtothist}
\end{figure}
%
%**********************************************************

The global parameters available in this study are still insufficient for the identification of distinguishing features between the three galaxy groups. Out of structural parameters like the concentration index, morphology, bulge-to-total ratio, (B/T), and effective surface brightness, only the total stellar and \hi\ masses (M$_{\mathrm{tot}}$ and M$_{\mathrm{\hi}}$) offer any differentiation between the three galaxy groups. \Fig{Mtothist} shows the A and B type galaxies are generally higher stellar and \hi\ mass systems, as expected, whereas the stochastic signature of C types is found in less massive environments.  The slight shift upward in total stellar mass (M$_{\mathrm{tot}}$) of A type galaxies suggests a more active history of accumulation of stellar mass via higher SFRs in the past. The matching offset in total \hi\ mass is however not substantial enough to cleanly delineate the A and B populations. It can be said that As simply have a slightly larger star formation fuel reservoir.
No distinction in bulge strength (B/T ratio) was observed between type A and B galaxies, though perhaps the activity within the bulge differentiates the two populations, rather than the bulge size itself.
Mechanisms such as central mass accretion or SF quenching may also explain the behaviors seen in type A and B galaxies.  In order to understand better the different mechanisms distinguishing these galaxy track types, the mean radial profiles of these three groups must now be examined in \sec{radial}.

%--------------------------RADIAL PROFILES----------------------------------%
\section{Radial Profiles}
\label{sec:radial}

%**********************************************************
%%Figure: Radial Profile
%
\begin{figure}%[b]
\centering
\includegraphics[width=0.33\textwidth]{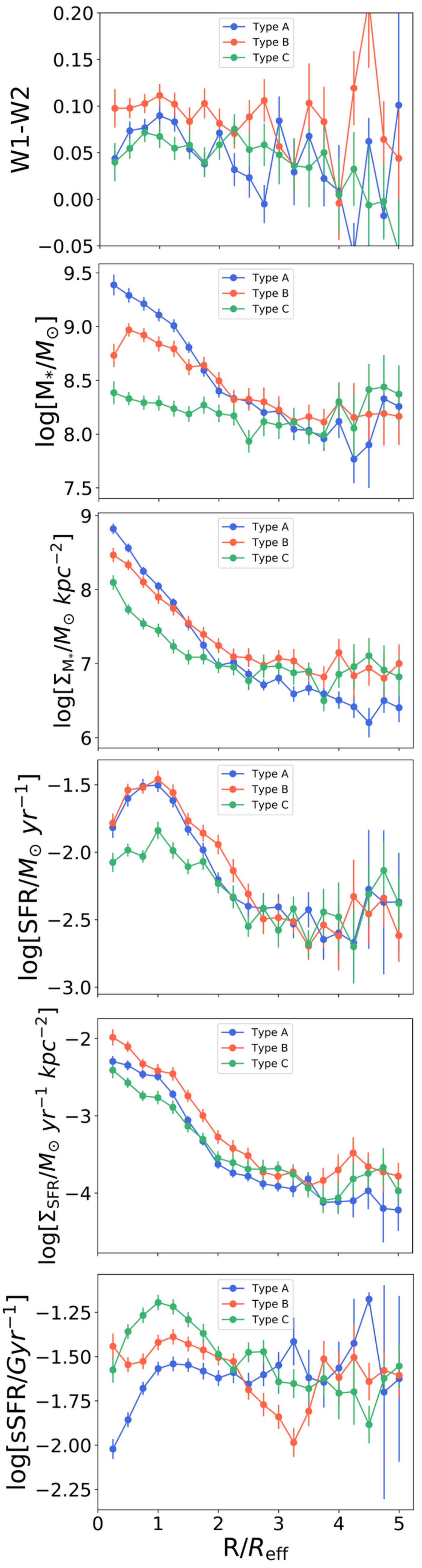}
\caption{Stacked radial profiles of $W1-W2$ color, \mstel, $\Sigma_{\mstel}$, SFR, $\Sigma_{\mathrm{SFR}}$ and sSFR for galaxies classified in the Type A, B, C galaxy groups (see \Fig{tracks} and \sec{scatter}). The \mstel\ used to calculate these profiles were derived from a constant $M_{*}/L_{W1}=0.5$, and SFRs used the \cite{calzetti07} transformation.
Radial values beyond $R/R_{\mathrm{eff}}=2.5$ become less robust, as fewer galaxies extend that far.}
\label{fig:radialprofiles}
\end{figure}
%
%**********************************************************

\Fig{tracks} shows the type A, B, and C tracks of individual galaxies across the local SFMS.  We now analyze their mean radial profiles, with a special focus on disentangling the type A and B populations.

\Fig{radialprofiles} shows radial profiles of $W1-W2$ color, \mstel, $\Sigma_{\mstel}$, SFR, $\Sigma_{\mathrm{SFR}}$, and sSFR that have been calculated by stacking profiles of galaxies from the same class (A, B, or C) and scaled to units of the \textit{W}1 effective radius ($R_{\mathrm{eff},W1}$). Note that the averages at larger radial values become statistically weaker, as fewer galaxies extend to such radii. We therefore place greater attention to regions within 2.5~$R_{\mathrm{eff}}$ as the majority of galaxies in our sample encompass that size.

In \sec{scatter} we posited that either the mass accretion history or SF quenching may separate these galaxy classes. 
Quenching behavior is generally associated with suppressed SFR relative to the stellar mass growth.
This can be defined qualitatively by the sSFR value (SFR/\mstel). High sSFR values may represent systems with active SF (high SFR) compared to the existing stellar mass, whereas a low sSFR may represent systems with a significant amount of \mstel\ and/or relatively low SF activity.

The average radial profiles of sSFR values across galaxies may reveal whether galaxies quench from the inside-out or outside-in, if at all. Inside-out quenching (\citealt{forbes14,gonzalez16,tacchella16b,belfiore17, abdurro18}) is dominant in higher stellar mass (\citealt{kimm09,perez13}) and early-type \citep{medling18} galaxies; potentially encouraged in barred galaxies \citep{abdurro17}. Outside-in quenching may be influenced by dense galaxy cluster environments (\citealt{schaefer17,medling18}) or by a lower total stellar mass \citep{perez13}. Typical star-forming sSFR values are higher than $10^{-2}$~Gyr$^{-1}$, whereas quenched environments have values less than $10^{-2}$~Gyr$^{-1}$. 
These limits are not strict, and still debated. For example, \cite{tacchella16b} found inside-out quenching, with `quenched' environments measuring log(sSFR)$\sim$-2, whereas \cite{medling18} identified galaxies with log(sSFR)$\sim$-3 to be quenched.
We use these thresholds as guidelines and search for `quenched' signatures in our sample, and whether these reflect a suppressed SFR or a higher central build-up of stellar mass.

The bottom panel of \Fig{radialprofiles} shows the sSFR profile of the type A, B, and C groups. The type A population displays a distinct depression of sSFR within the inner parts of the galaxy, compared to type B, and even type C groups.  It appears SF activity in class A systems is delayed relative to the amount of stellar mass that has accumulated at the center. Beyond 1-2~$R_{\mathrm{eff}}$, the quenching signature dissipates.  Similar signatures of inside-out quenching are reported in Fig.~7 of \cite{tacchella16b} and Fig.~3 of \cite{belfiore17}.  The latter comparison indicates that the onset of quenching arises first in high stellar mass galaxies. The galaxy is then at a transition point from the  star-forming sequence on its way to becoming fully quenched.

Fully quenched galaxies are predominantly early-types \citep{gonzalez16,medling18}, and since our sample is dominated by spirals and low mass elipticals, significantly quenched systems should be less prevalent. However, the type A population seems to feature a transition point for the onset of quenching for late-type galaxies.
This signature may very well arise from the presence of a bulge. Indeed, \cite{forbes14} suggest that a bulge can quench the inner regions. Type A galaxies will likely reach higher total stellar masses, and thus mos likely host a bulge as well \citep{wang17}. Similarly, \cite{ellison18} propose that central starbursts produce a large central mass build up, leading to inside-out quenching.

In \Fig{radialprofiles}, the SFR profiles of the type A and B populations track each other almost exactly, both with a central dip perhaps indicating a weakly star forming bulge. This is contrary to expectations from the sSFR profiles, calling for higher SFR measurements from the type B population. This draws attention to the stellar mass profiles. Here, the type A population has significantly higher stellar mass values near the center relative to the type B population. This demonstrates that the `quenched' sSFR signature is in fact the result of a larger accumulation of central mass (and perhaps a longer SFH) in the type A galaxies, while SFR is still ongoing. Another possible factor contributing to the lower \mstel\ presence at the center of type Bs may be stellar migration in disks. It is possible that in this galaxy population, mechanisms pushing stars outward in the disk are more dominant, as seen by \cite{frankel18}.

The \mstel- and SFR-density ($\Sigma_{\mstel}$, $\Sigma_{\mathrm{SFR}}$) profiles reveal additional clues, along with the pure \mstel and SFR profiles.
The density normalization in the $\Sigma_{\mstel}$ and $\Sigma_{\mathrm{SFR}}$ measurements shows a decreasing gradient outwards radially in both measurements and no significant suppression of central SF.  While the pure \mstel\ and SFR profiles highlight nuances to separate these populations, the sSFR profiles can confirm or invalidate the quenching feature. 
Indeed, the sSFR profile confirms a `transitional' quenching feature in type As, with a large accumulation of mass indicating the now active SF will likely soon slow down.
The density profiles also hint at the size variations amongst the three galaxy populations. Since the density measurements normalize both the \mstel\ and SFR values, it suggests that the type A galaxies are generally larger.
Therefore the Type C galaxies are likely smaller in size. This is consistent with \Fig{Mtothist} and the distribution of total stellar mass associated with each population.

The $W1-W2$ color is sensitive to the relative dust content in the region of study, hinting at the remaining fuel for SF. As discussed in \sec{stellm}, redder colors suggest a larger dust content, whereas blue colors suggest emission by old stellar populations with a minimal contribution from dust. In the $W1-W2$ profile, type B galaxies have, on average, redder $W1-W2$ values, thus hinting at a significant dust fraction. This would be consistent with continually active SF regions. In contrast, the bluer $W1-W2$ color of type A galaxies suggests they contain an older stellar population. While the SFR profiles of type A and B galaxies may be comparable, the potentially higher dust content implied by $W1-W2$ suggests type Bs have a deeper fuel reservoir (dust and gas) to supply ongoing SF for a longer period than type As. This seems consistent with inferences based on the stellar mass profiles, where the type A population has accumulated a considerable amount of stellar mass at the center of the galaxies, and will likely exhaust its fuel before the type B systems.

Conversely, the type C population has, on average, bluer $W1-W2$ colors suggesting lower dust densities, as expected for galaxies of lower total stellar mass \citep{dalcanton04}. Lower SFRs for these galaxies scales with lower dust levels in this population, as gas (and dust) would be necessary to fuel the SF activity. sSFRs across type C galaxies are higher, but this is driven by the drastically lower \mstel\ values. The sSFR curve also demonstrates a slight inner quenching signature, though the order of magnitude is still much higher than that of the type A curve, and is likely representative of the sparse stellar population in smaller galaxies, rather than truly quenched environments. 

Overall, \Fig{radialprofiles} shows that the onset of inside-out quenching signatures exists in some spiral galaxies, here labeled as type A galaxies.  We surmise that the driving factor is not a reduced SFR, but rather a larger accumulation of stellar central mass. This may arise from one or a combination of three scenarios: (1) the galaxy is older, and its SFH is therefore more extended; (2) the galaxy has accreted more mass from external sources in order to fuel continual central SF, or (3) a bar funnels fuel for SF toward the center of the galaxy.  The real answer is likely a combination of these three factors, though the first two are difficult to verify observationally. These galaxies may soon transition into the `green valley' range of the SFMS once they have completely exhausted their SF fuel. Finally, type B and C galaxies behave as expected for high and low-mass galaxies within the SFMS.

%-----------------------------------------------------CONCLUSION------------------------------------------------%
\section{Summary \& Conclusions}
\label{sec:summary}

We have presented a study of the local and global SFMS and established their comparable behaviors. Using our preferred \mstel\ and SFR conversions, we find the fit parameters for the relation $\log{(\mathrm{SFR})} = a\log{(\mstel)} +b$ to be $a=1.03$, $b=-10.63$, and $\sigma=0.30$, for the spatially resolved SFMS (see \Table{fitstable}).

Our study of the SFMS considers radial binning scales down to about 50~pc. We have determined the ideal resolution scale to be 500~pc or larger in order to properly encapsulate the physics inherent in the applied transformations.

Our study of the local SFMS uncovered three populations characterized by their radial tracks across the SFMS (\Fig{tracks}). These classes, informally referred to as type A, B, and C galaxies, are characterized as follows:
\renewcommand{\labelitemi}{{\boldmath$\cdot$}}
\begin{itemize}
\item Type A galaxies show a sharp rise in SFR at a relatively constant \mstel\ in their inner regions, then decline with decreasing \mstel\ and SFR values along the SFMS slope. 
\item Type B galaxies show no sharp rise in SFR at the center, but parallel the SFMS at all radial points also, with finely tuned decreasing \mstel\ and SFR values. 
\item Type C galaxies, for which all local log($\Sigma_{\mstel}$)$\sim$7.5 show no correlation with SFMS within a galaxy, but \mstel\ and SFR measurements are more sporadic.
\end{itemize}

Signatures of inside-out quenching are seen in type A galaxies. It does not appear to arise from suppressed SFR, but rather a larger build up of stellar mass at the center. We propose that this is likely encouraged by a combination of (1) galaxies being older and therefore having a longer SFH, (2) galaxies having accreted more mass from external sources to fuel continual SF, or (3) the presence of a bar feeding fuel for SF towards the center.
These galaxies may be at a transitionally quenched phase, and as they exhaust their SF fuel, slide into the `green valley' range of the SFMS (as suggested by \citealt{belfiore17}), perhaps encouraged by central starbursts (as suggested by \citealt{ellison18}).

The scatter in the local SFMS appears to be driven by different mechanisms above and below a critical stellar mass density of log($\Sigma_{\mstel}$)$\sim$7.5, yet the same scatter amplitude persists at all stellar mass ranges (in the log$-$log space of the SFMS). 
Above local measurements of log($\Sigma_{\mstel}$)$\sim$7.5, type A+B galaxies slide along the SFMS slope, but are vertically offset from each other. The scatter in this stellar mass density regime is likely dominated by systematic differences in galaxies and their varying global environments. 
Below local measurements of log($\Sigma_{\mstel}$)$\sim$7.5 type C galaxies and even the tails of the type A+B galaxies no longer follow the SFMS, but measurements are more sporadic. Scatter is likely dominated by stochastic sampling of SF behavior and environments at such small stellar masses.

The SFMS residual analysis (see also \citealt{dutton07,brennan17}) does not reveal a noticeable correlation with any single structural parameter. 
In particular, morphology and the \hi\ gas fraction do not appear to affect the scatter of the global or local SFMS, as suggested by other studies \citep{gonzalez16,saintonge16,gonzalez17}.

%-----------------------------------------------------ACKNOWLEDGMENTS------------------------------------------------%
\section{Acknowledgments}

\noindent

We thank Aaron Dutton and Andrea Macci\`o for thoughtful discussion enhancing our study.
We thank Dennis Zaritsky and Sharon Meidt for insightful conversations about mid-infrared colors and mass-to-light ratios.  
We thank Richard Tuffs for useful suggestions based on an earlier draft. 
S.C. acknowledges support from the NSERC of Canada through a Research Discovery Grant.

\bibliography{HallChristine_SFMSmanuscript.bbl}

\section*{APPENDIX A: The Radial Extent of Galaxies in Various Bands}

\begin{figure*}[t!]
%\captionsetup{width=0.9\textwidth}
\centering
\subfigure[]{\includegraphics[width=0.45\textwidth]{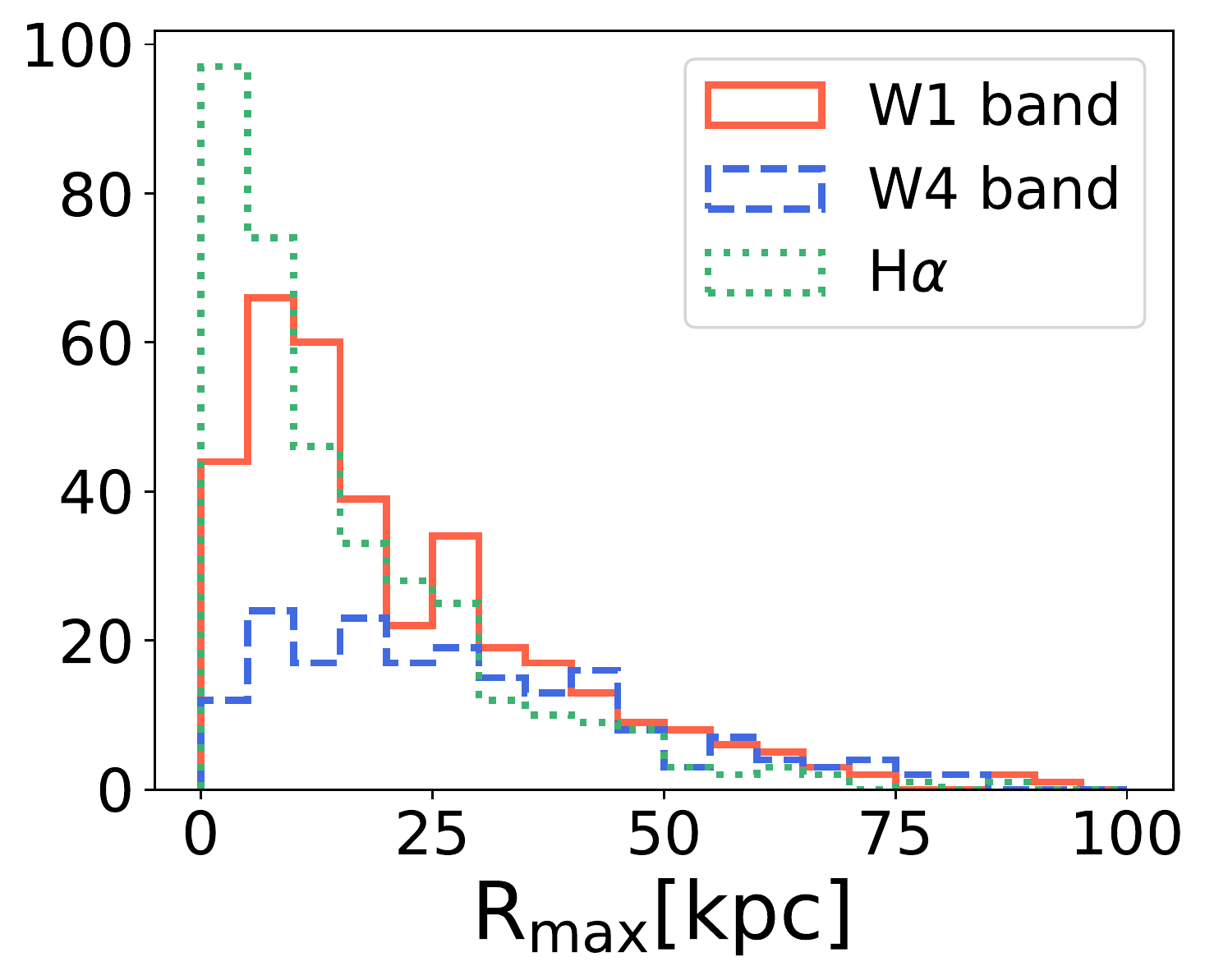}}
\hfill
\subfigure[]{\includegraphics[width=0.45\textwidth]{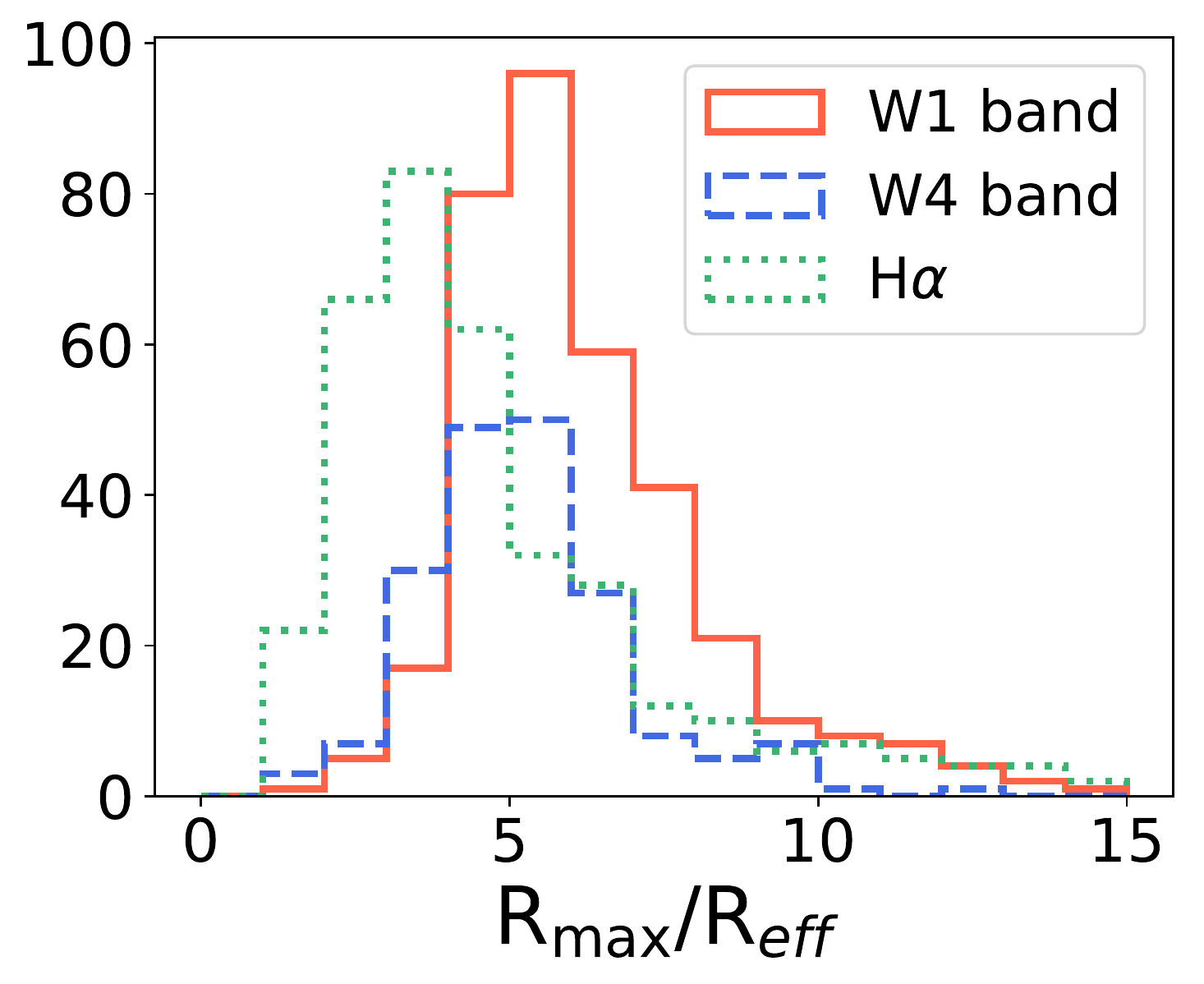}}
\caption{The radial extent (R$_{\mathrm{max}}$) of H$\alpha$ emission and  \textit{W}1 \& \textit{W}4 band profiles for all galaxies scaled by (a) kiloparsecs (kpc) and (b) effective radius ($R_{\mathrm{eff}}$).}
\label{fig:Rhist}
\end{figure*}

A major advantage of the \textit{WISE} and SINGG dataset is the large extent of the radial profiles. Other surveys such as MaNGA \citep{albar17} or CALIFA \citep{sanchez16} limit the majority of their galaxies to 1.5~$R_{\mathrm{eff}}$ or 4.2~$R_{\mathrm{eff}}$, respectively. \Fig{Rhist} displays the radial extent of all galaxies in our sample, specifically for H$\alpha$ emission from SINGG and \textit{W}1 and \textit{W}4 band emission from \textit{WISE}. 
We rely predominantly on the H$\alpha$ conversion (\eq{KSFR}), or on the scenario of \eq{CSFR} which uses predominantly H$\alpha$ and a smaller contribution of \textit{W}4; regions without \textit{W}4 emission will represent where there is no re-radiated dust, and SFR is traced by the H$\alpha$ component.

%**********************************************************
%Figure: Data extension
%
%
%**********************************************************

\section*{APPENDIX B: Recasting the \mstel\ Transformation.}

As discussed in \sec{stellm}, the \mstel\ transformation by \Esk\ (\eq{EMstar}) is given in a very different form than the \Cla\ and \Me\ transformations. We recast this transformation into the language of the other two forms, as given by \eq{EMstar2}. Here we outline the procedure to recast the transformation.

First we rearrange the original formula to accommodate distances in pc (no longer Mpc):
\small
\begin{align*}
M^{*}_{\mathrm{Eskew}} &= 10^{5.65}~F^{2.85}_{W1}(\mathrm{Jy})~F^{-1.85}_{W2}(\mathrm{Jy})~\bigg[\frac{D(\mathrm{Mpc})}{0.05}\bigg]^{2}\\
 &= 10^{5.65}~F^{2.85}_{W1}(\mathrm{Jy})~F^{-1.85}_{W2}(\mathrm{Jy})~\bigg[20\times\frac{D(\mathrm{pc})}{10^{6}\mathrm{(pc/Mpc)}}\bigg]^{2}\\
 &= 10^{-3.75}~F^{2.85}_{W1}(\mathrm{Jy})~F^{-1.85}_{W2}(\mathrm{Jy})~D^{2}(\mathrm{pc})
\end{align*}
\normalsize

Next, in order to convert the fluxes in Jy to magnitudes, we utilize the ``Zero Point Magnitudes'' ($F_{\nu}(iso)$) reported by \cite{jarrett17} (the calibration detailed by \citealt{jarrett11}); given a flux density in Jy ($S_{\nu}$), the magnitude is determined by $m_{\nu}=-2.5\times \log_{10}[S_{\nu}/F_{\nu}(iso)]$. For the \textit{W}1 and \textit{W}2 bands; $F_{W1}(iso)=309.68$ Jy and $F_{W2}(iso)=170.66$ Jy, which results in:

\small
\begin{align*}
M^{*}_{\mathrm{Eskew}} = & 10^{-3.75}~ (309.68\times10^{-0.4~m_{W1}})^{2.85}\\
									&   \times (170.66\times10^{-0.4~m_{W2}})^{-1.85}~D^{2}(\mathrm{pc}) \\
 									= & 10^{-3.75}~309.68^{2.85}~170.66^{-1.85} 10^{-1.14~m_{W1}}\\
									&   \times 10^{0.74~m_{W2}}~D^{2}(\mathrm{pc}) \\
 \log(M^{*}_{\mathrm{Eskew}}) = & -0.79-1.14m_{W1}+0.74m_{W2}+2\log(D(\mathrm{pc}))
\end{align*}
\normalsize

In order to extract the mass-to-light ratio, we must pull log($L_{\odot,W1}$) from the right hand side of the above equation. We know that log$(L_{\odot,W1})=0.4M_{\odot,W1}-0.4M=0.4M_{\odot,W1}-0.4m_{W1}+2\log(D(pc))-2$. (\cite{randri14} and \cite{jarrett13} report the value $M_{\odot,W1}=3.24$). Therefore we can write the above equation as:

\small
\begin{align*}
 \log(M^{*}_{\mathrm{Eskew}}) = & -0.79-0.74m_{W1}-0.4m_{W1}\\
 												&	+0.74m_{W2}+2\log(D(\mathrm{pc}))-2+2\\
 												&  +0.4M_{\odot,W1}-0.4M_{\odot,W1} \\
 \log(M^{*}_{\mathrm{Eskew}}) = & -0.79+2-0.4M_{\odot,W1}-0.74m_{W1}\\
 												&  +0.74m_{W2}+\big(0.4M_{\odot,W1}-0.4m_{W1}\\
 												&  +2\log(D(\mathrm{pc}))-2\big) \\
  \log(M^{*}_{\mathrm{Eskew}}) = & -0.074-0.74\big(m_{W1}-m_{W2}\big) +\log(L_{\odot}) \\
  \log(M^{*}_{\mathrm{Eskew}})-\log(L_{\odot}) = & -0.074-0.74\big(W1-W2\big) \\
  \log(M^{*}_{\mathrm{Eskew}}/L_{\odot}) = & -0.074-0.74\big(W1-W2\big) \\
\end{align*}
\normalsize

With the stellar mass equation in this form, we can directly compare each method, as seen in \sec{stellm}.

\section*{APPENDIX C: Subsidiary SFMS Fits.}

In our main analysis, we have opted for a constant $M_{*}/L_{W1}=0.5$ to calculate the stellar masses.
We now include the SFMS fits produced by the \Esk\ and \Cla\ in Table~\ref{app:fitstable}, for completeness.
Comparing the various $a$, $b$, and $\sigma$ values, we find that the choice of \mstel\ transformation
does not significantly impact the fit of the SFMS.

%**********************************************************
%Table: Main Sequence Fits
%
\begin{table*}[h]
\centering
\caption{Subsidiary Star Formation Main Sequence Fit Parameters}
\begin{tabular}{ | c | c c c c c | }
\hline
 & \mstel\ Transformation & SFR Transformation & Slope $a$ & Zero Point $b$ & Standard Deviation $\sigma$ \\ [0.5ex]
\hline\hline
Global (G):	& \citealt{eskew12}  				& \citealt{kennicutt98b}  				& 0.79 	& -8.43  	& 0.31 \\
Local (L):  	& 	(Eq.~\ref{eq:EMstar})		& 	(Eq.~\ref{eq:KSFR})        			& 1.04 	& -10.97 	& 0.37 \\
\hline
G:          	& 													& \citealt{cluver17} \textit{W}3 			& 1.13 	& -11.72 	& 0.21 \\
L:          		& 													& (Eq.~\ref{eq:ClSFR1})    					& 1.02 	& -10.28 	& 0.27 \\
\hline
G:          	& 													& \citealt{cluver17} \textit{W}4 			& 1.01 	& -10.58  	& 0.29 \\
L:          		& 													& 	(Eq.~\ref{eq:ClSFR2})					& 0.98 	& -10.12 	& 0.36 \\
\hline
G:          	& 													& \citealt{calzetti07}    				& 0.83 	& -8.96  	& 0.30 \\
L:          		& 													& (Eq.~\ref{eq:CSFR})  				& 1.02 	& -10.74 	& 0.31 \\
\hline
G:          	& 													& \citealt{davies16}                    		& 0.79 	& -8.25  	& 0.31 \\
L:          		& 													& 	H$\alpha$, old							& 1.04 	& -10.79 	& 0.37 \\
\hline
G:          	& 													& \citealt{davies16}                    	& 0.45 	& -4.43  	& 0.21 \\
L:          		& 													& 	H$\alpha$, new						& 0.72 	& -7.35  	& 0.30 \\
\hline
G:         	 	& 													& \citealt{davies16}              		& 0.83 	& -9.00  	& 0.18 \\
L:          		& 													& 	\textit{W}3, new 					& 0.82 	& -8.79  	& 0.21 \\
\hline
G:          	& 													& \citealt{davies16}                     & 0.51 	& -5.39  	& 0.19 \\
L:          		& 													& 	\textit{W}4, new					& 0.64 	& -6.91  	& 0.25 \\
\hline\hline
G:          	& \citealt{cluver14}  				& \citealt{kennicutt98b}  		& 0.78 	& -8.21  	& 0.35 \\
L:          		& (Eq.~\ref{eq:CMstar})			& 		(Eq.~\ref{eq:KSFR})			& 1.02 	& -10.61 	& 0.39 \\
\hline
G:          	& 													& \citealt{cluver17} \textit{W}3 		& 1.14 	& -11.60 	& 0.27 \\
L:          		& 													& (Eq.~\ref{eq:ClSFR1})					& 0.80 	& -8.48 	& 0.27 \\
\hline
G:          	& 													& \citealt{cluver17} \textit{W}4 		& 1.01 	& -10.37  	& 0.34 \\
L:          		& 													& (Eq.~\ref{eq:ClSFR2})				& 0.62 	& -6.61 	& 0.36 \\
\hline
G:          	& 													& \citealt{calzetti07}      			& 0.83 	& -8.74  	& 0.34 \\
L:          		& 													& (Eq.~\ref{eq:CSFR})				& 1.10 	& -11.04 	& 0.33 \\
\hline
G:          	& 													& \citealt{davies16}                    		& 0.78 	& -8.03  	& 0.35 \\
L:          		& 													& 	H$\alpha$, old							& 1.02 	& -10.42 	& 0.39 \\
\hline
G:          	& 													& \citealt{davies16}                    	& 0.44 	& -4.26  	& 0.23 \\
L:          		& 													& H$\alpha$, new						& 0.69	& -7.02  	& 0.27 \\
\hline
G:         	 	&													& \citealt{davies16} 		         	& 0.83 	& -8.85  	& 0.23 \\
L:          		& 													& \textit{W}3, new    					& 0.80 	& -8.48  	& 0.23 \\
\hline
G:          	& 													& \citealt{davies16}					& 0.50 	& -5.19  	& 0.22 \\
L:          		& 													& 	\textit{W}4, new					& 0.62 	& -6.61  	& 0.24 \\
\hline
\end{tabular}
\begin{tablenotes}
\item \textbf{Notes.} Numerical values for the global and local SFMS fits for the alternative stellar mass estimate methods; \cite{eskew12} and \cite{cluver14}. Fits are applied to the global $\log{(\mathrm{SFR})} = a\log{(\mstel)} +b$ and local $\log{(\Sigma_{\mathrm{SFR}})} = a\log{(\Sigma_{\mstel})} +b$ relations.
\end{tablenotes}
\label{app:fitstable}
\end{table*}
%
%**********************************************************

\section*{APPENDIX D: Subsidiary Analysis of Parameters Not Influencing the SFMS Scatter.}

\begin{figure*}
\centering
\subfigure[]{\includegraphics[width=5.4cm]{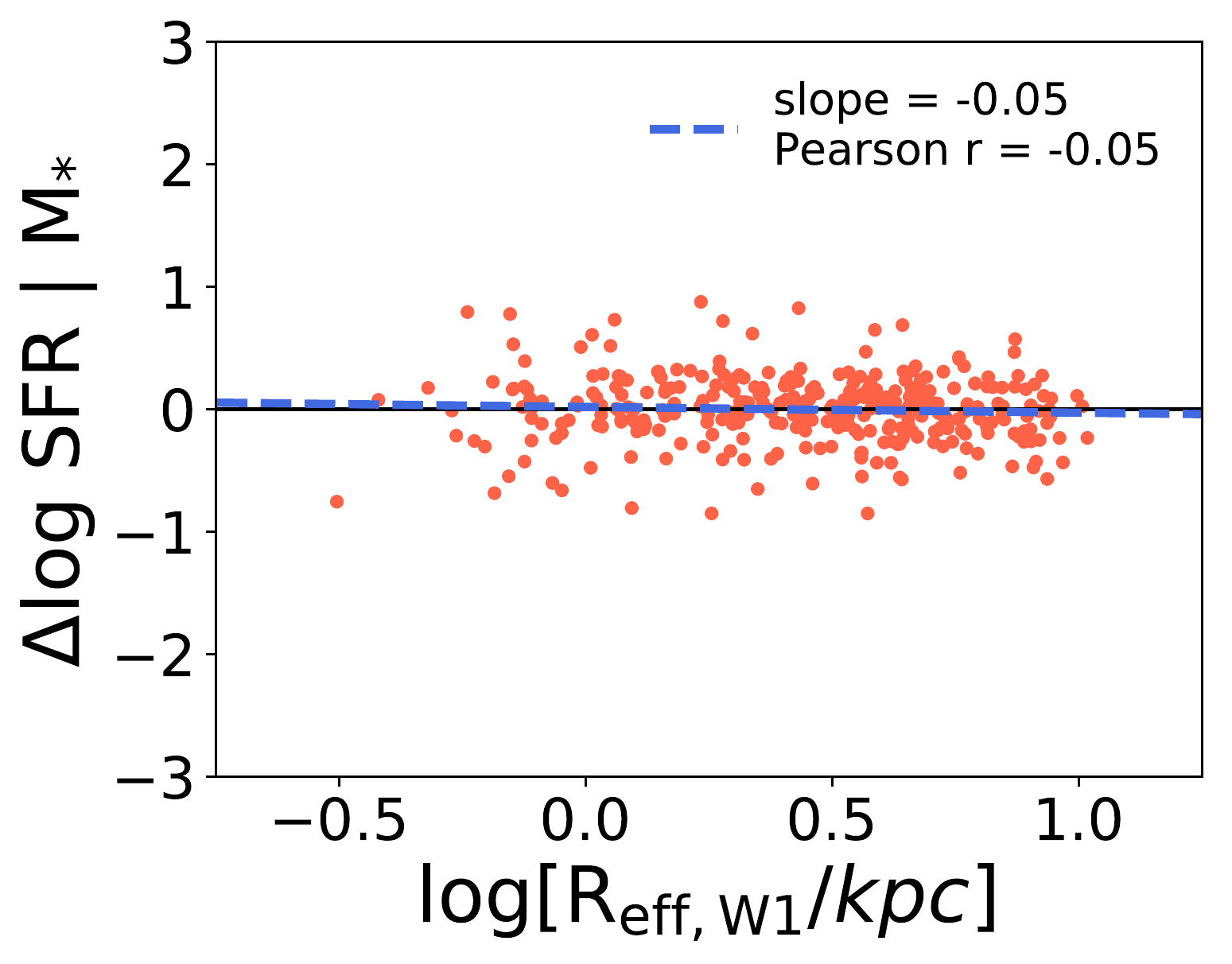}}
\hfill
\subfigure[]{\includegraphics[width=5.4cm]{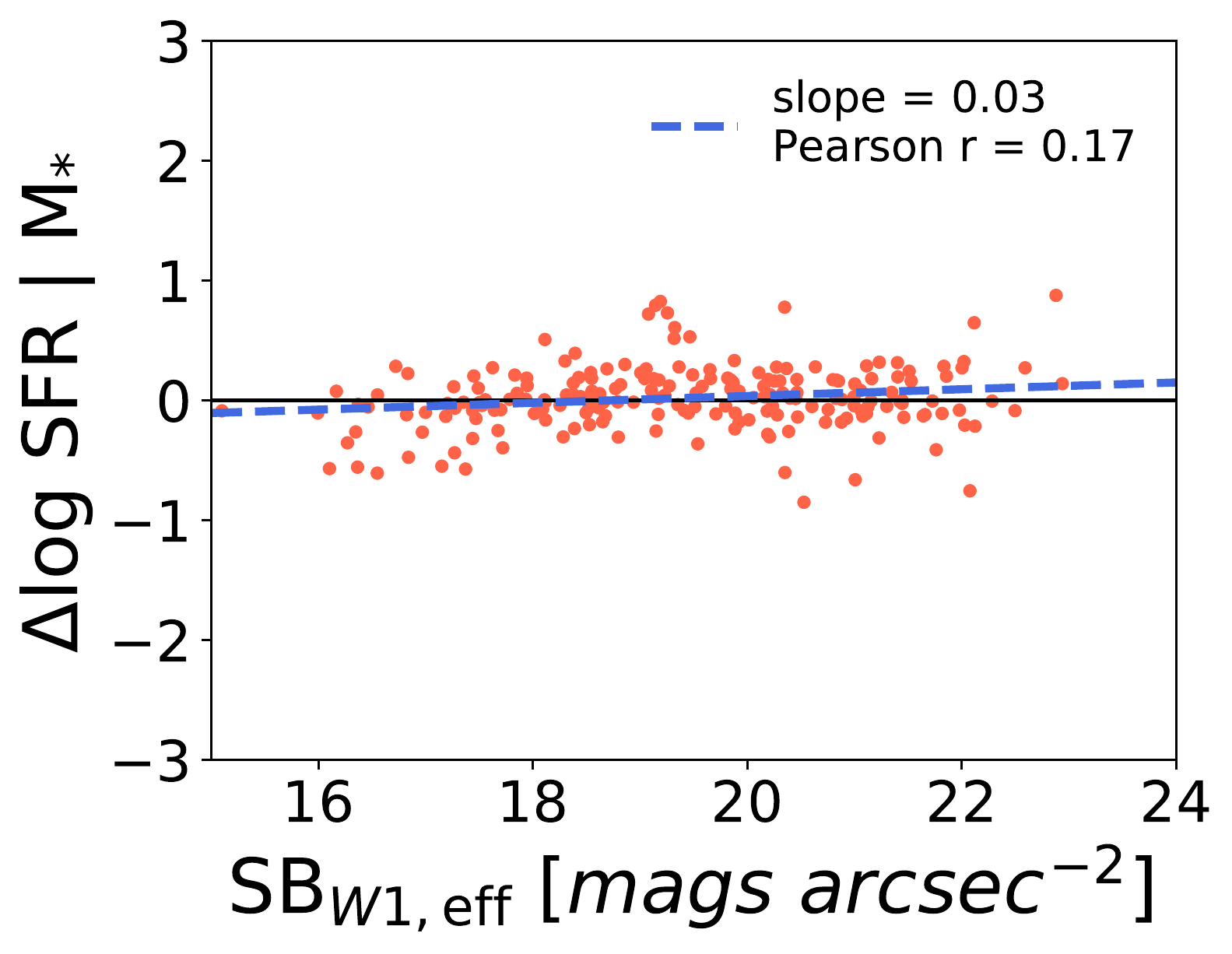}}
\hfill
\subfigure[]{\includegraphics[width=5.4cm]{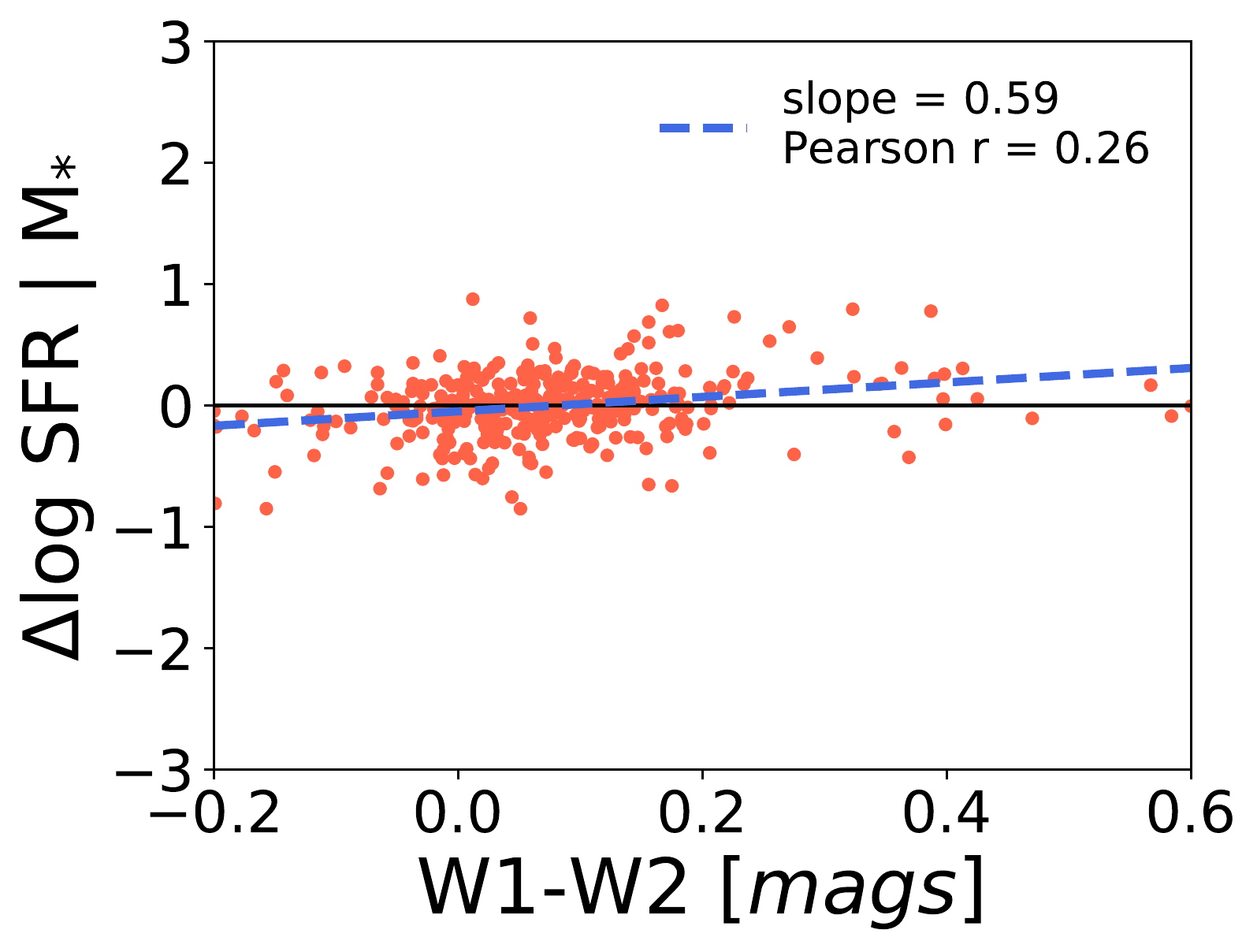}}
\vfill
\subfigure[]{\includegraphics[width=5.4cm]{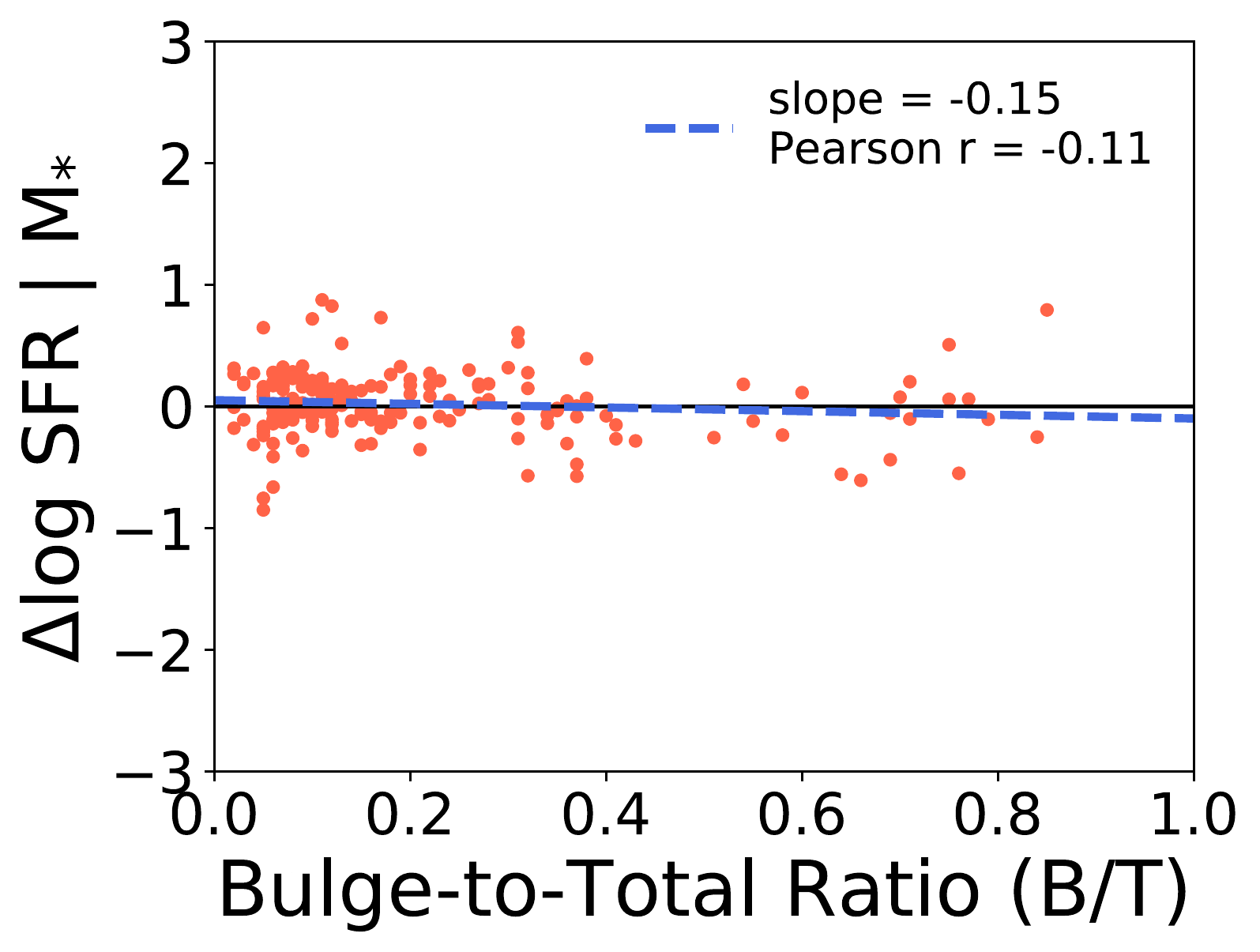}}
\hfill
\subfigure[]{\includegraphics[width=5.4cm]{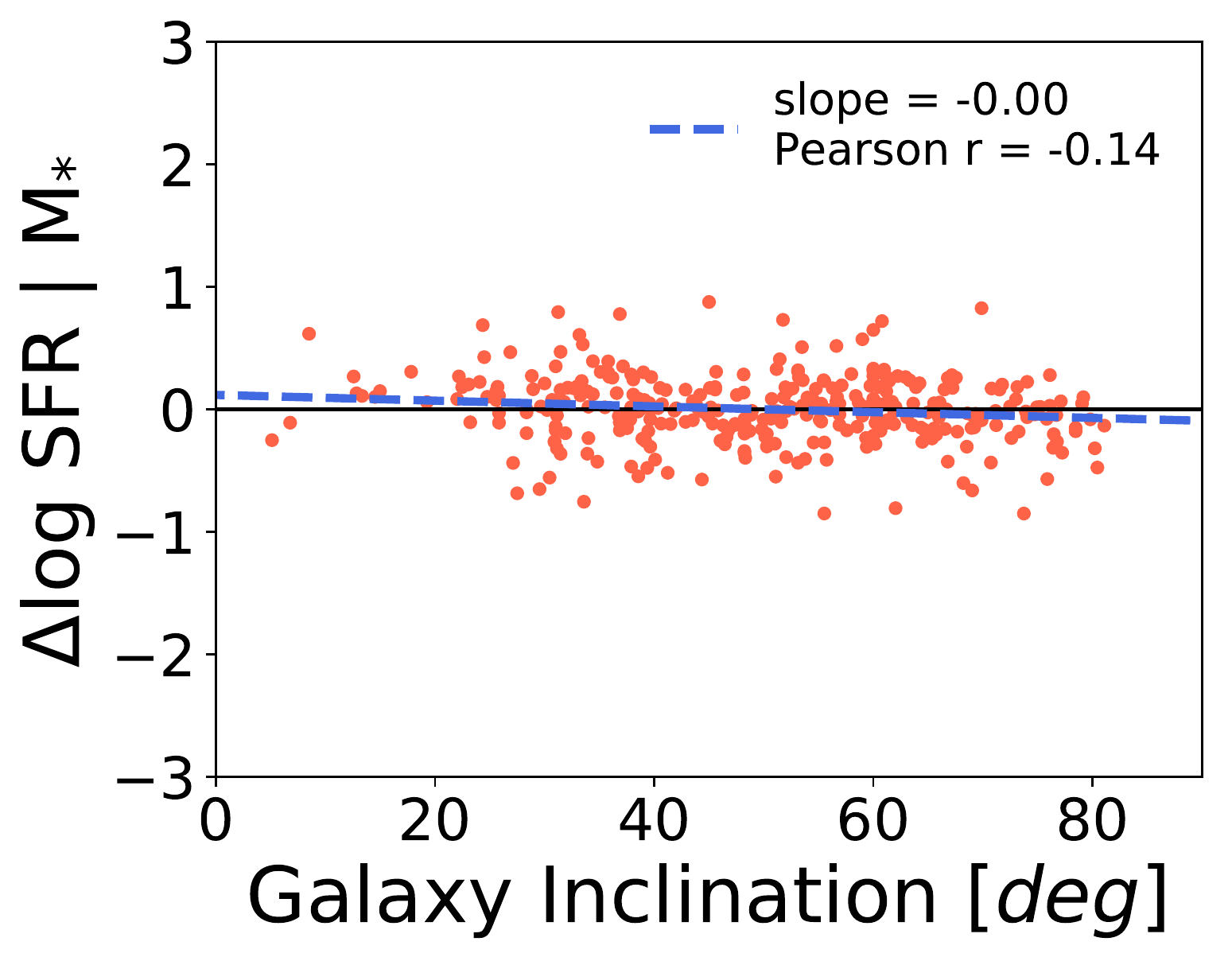}}
\hfill
\subfigure[]{\includegraphics[width=5.4cm]{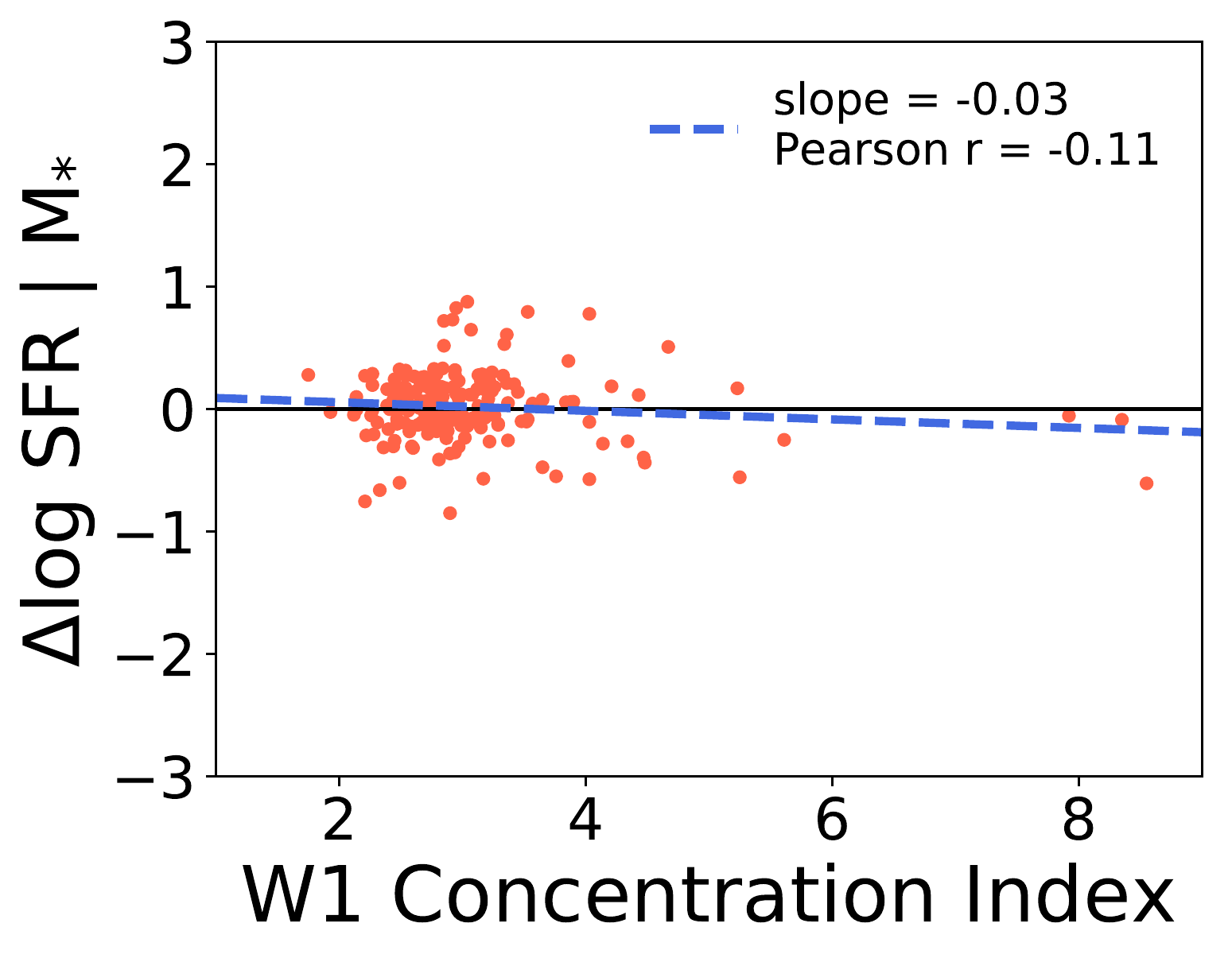}}
\vfill
\subfigure[]{\includegraphics[width=5.4cm]{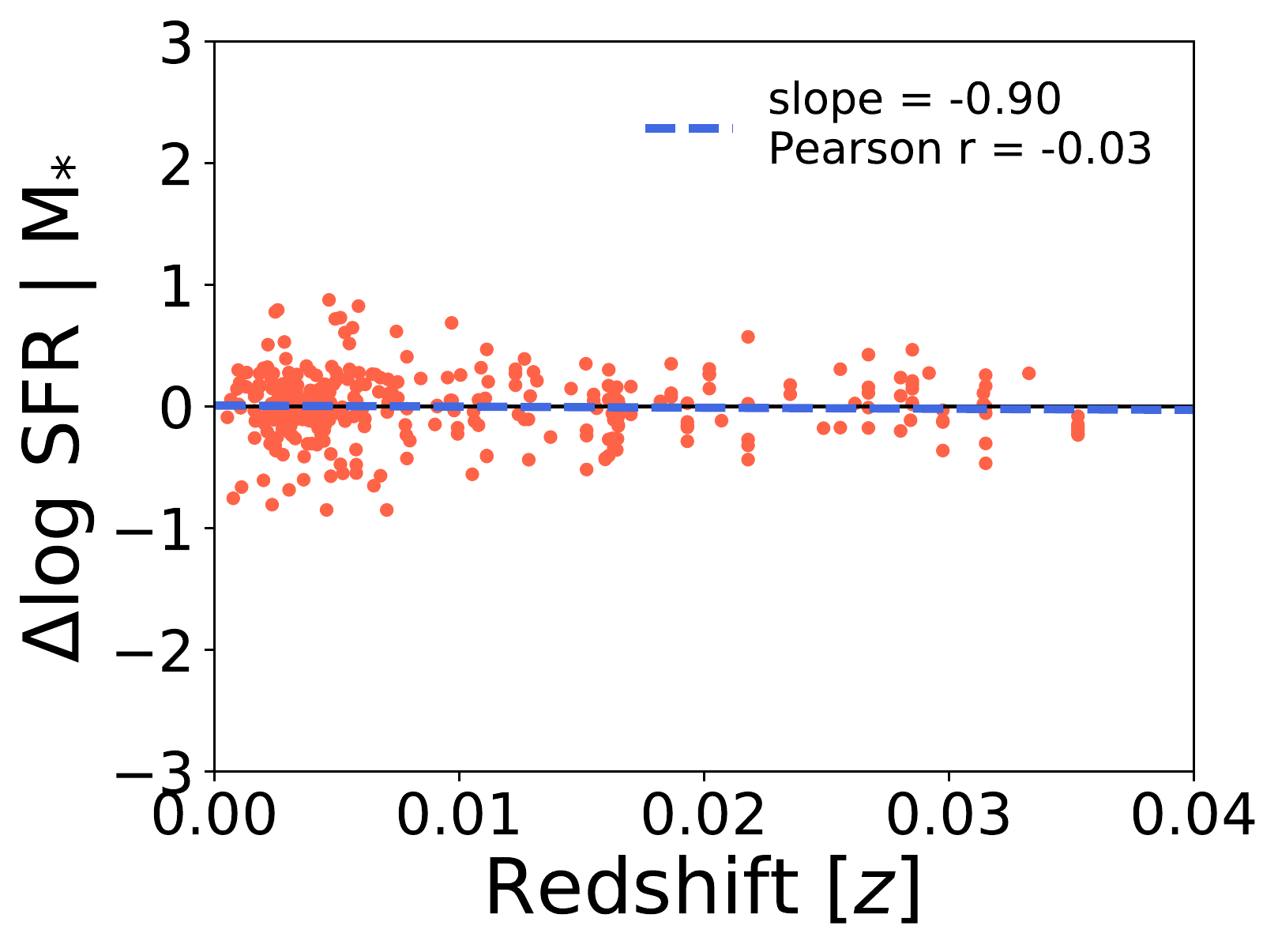}}
\hfill
\subfigure[]{\includegraphics[width=5.4cm]{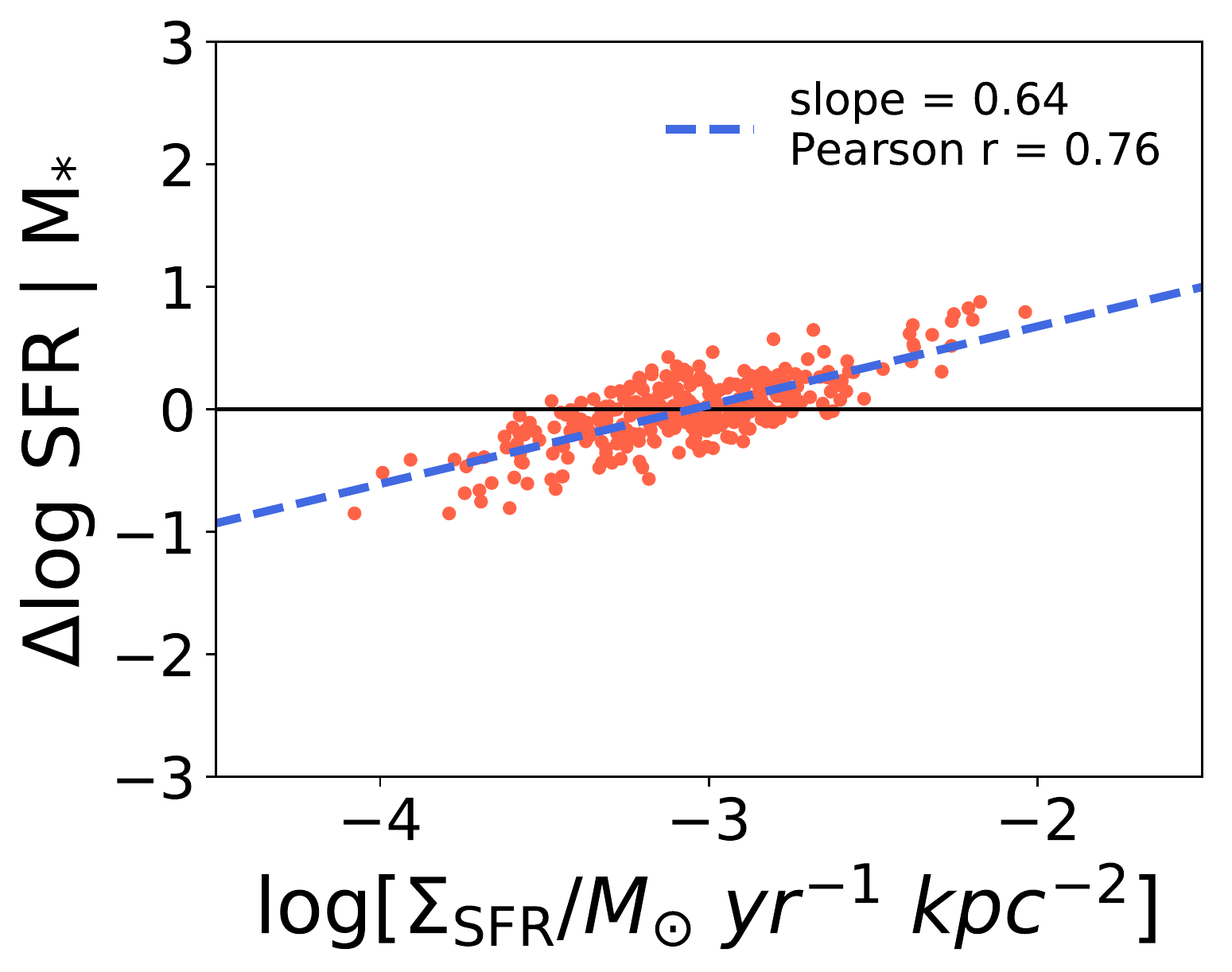}}
\hfill
\subfigure[]{\includegraphics[width=5.4cm]{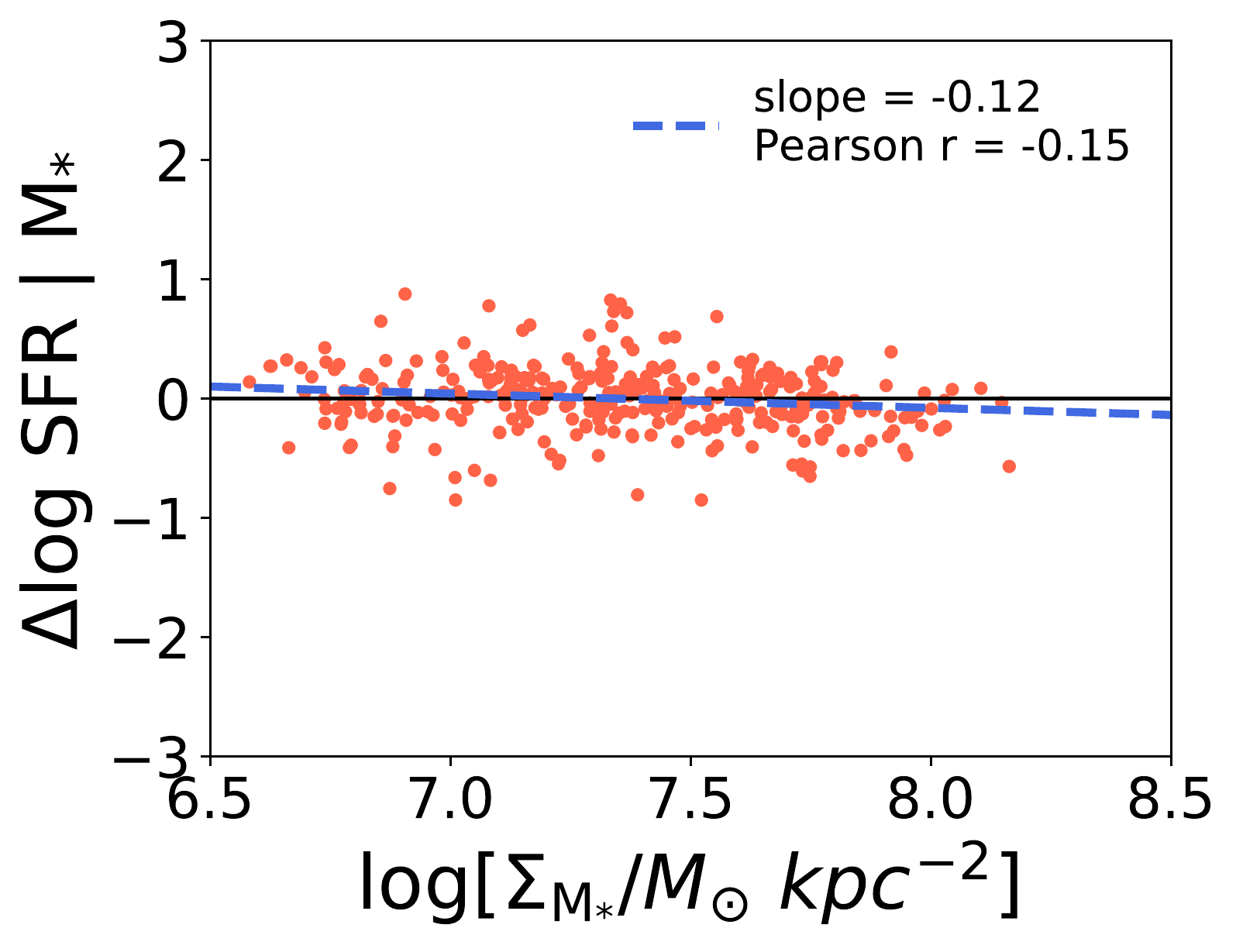}}
\caption{Vertical distance from global SFMS fit ($\Delta$log(SFR)) at a fixed \mstel\ against various parameters. The black line is the linear fit to the SFMS, the blue line is the fit the residuals corresponding to each global parameter. The correlation coefficient is given by Pearson r.}
\label{app:globresiduals}
\end{figure*}

We include here in Fig~\ref{app:globresiduals} subsidiary global residual plots that did not indicate any significant correlation of the corresponding parameter to the scatter of the SFMS (see analysis in \sec{scatter}). 

%**********************************************************
%Figure: Parameter residuals GLOBAL
%
%
%**********************************************************

\end{document}